\documentclass[prc,twocolumn,showpacs]{revtex4-1}

\usepackage{enumerate}

\usepackage{booktabs}
\usepackage{color}
\usepackage{graphicx}

\usepackage{amsmath}
\usepackage{amssymb}
\usepackage{times}
\usepackage{multirow}

\def\bvec#1{\mbox{\boldmath $#1$}}

\newcommand{\del}[2]{\frac{\partial #1}{\partial #2}}

\newcommand{\bra}{\langle}
\newcommand{\ket}{\rangle}

\usepackage{ascmac}
\usepackage{fancybox}

\newcommand{\beq}{\begin{equation}}
\newcommand{\eeq}{\end{equation}}
\newcommand{\bea}{\begin{eqnarray}}
\newcommand{\eea}{\end{eqnarray}}

\def\fun#1#2{\lower3.6pt\vbox{\baselineskip0pt\lineskip.9pt
 \ialign{$\mathsurround=0pt#1\hfil##\hfil$\crcr#2\crcr\sim\crcr}}}


\begin{document}

\title{
Eigenphase shift decomposition of the RPA strength function based on the Jost-RPA method
}

\author{K. Mizuyama$^{1,2}$, T. Dieu Thuy$^{3}$, T. V. Nhan Hao$^{3,4}$}

\affiliation{
  \textsuperscript{1}
  Institute of Research and Development, Duy Tan University,
  Da Nang 550000, Vietnam
  \\
  \textsuperscript{2}
  Faculty of Natural Sciences,  Duy Tan University, Da Nang 550000, Vietnam
  \\
  \textsuperscript{3}
  Faculty of Physics, University of Education, Hue University,
  34 Le Loi Street, Hue City, Vietnam \\
  \textsuperscript{4} Center for Theoretical and Computational Physics,
  University of Education, Hue University, 34 Le Loi Street, Hue City, Vietnam
}
\email{corresponding author: tvnhao@hueuni.edu.vn}

\date{\today}

\begin{abstract}
  The S-matrix which satisfies the unitarity, giving the poles as RPA excited states,
  is derived using the extended Jost function within the framework of the RPA theory. 
  An analysis on the correspondence between the component decomposition of the RPA strength function
  by the eigenphase shift obtained by diagonalisation of the S-matrix and the S- and K-matrix poles
  was performed in the calculation of the $^{16}$O quadrupole excitations.
  The results show the possibility that the states defined by the eigenphase shift can be expressed
  as RPA-excited eigenstates corresponding to the S-matrix poles in the continuum region. 
\end{abstract}

\maketitle

\section{Introduction}

In nuclear physics, which is a finite quantum many-body system, there are mainly two phenomena
called ``resonance''. One is the ``resonance'' that occurs in nuclear reactions and the other is
the ``resonance'' that refers to a collective excitation of the nuclear structure ({\it e.g.}
the giant resonance). Originally, ``resonance'' was referred to a vibration phenomenon in classical mechanics
in which the amplitude of a vibrating body increases when it is stimulated by an external vibration
equal to its eigenfrequency.
The term ``collective vibration mode'' in nuclear structures is sometimes called ``resonance'', {\it e.g.}
the giant resonance, because it is considered to be analogous to ``resonance'' in classical mechanics.
Such quantum states are generally represented by a linear combination of several different quantum
states and are known as metastable states.
In scattering theory, a resonance as a metastable state appears as a peak with width in the cross
section, etc., and the peak energy and width are considered to be given by the real and imaginary
parts of the poles of the S-matrix on the complex energy plane, respectively.
This is due to the fact that the resonance formula, such as the so-called Breit-Wigner type, has
been derived from models which reproduce the resonance state, and it has been confirmed that the
resonance energy and width are given by the real and imaginary parts of the S-matrix poles. 
However, it is not strictly guaranteed that the real part of the poles of the S-matrix coincides
with the resonance peak of the cross section, etc., due to the properties of the S-matrix as a
multivalued function of energy~\cite{jost-class}. Moreover, due to the special quantum transition to continuum
states and interference effects such as the Fano effect~\cite{jost-fano}, the cross section can become concave or
asymmetric shape near the real part of the poles of the S-matrix. 
Another way of determining the resonance state (resonance energy) is to use the K-matrix; it is known
that the poles of the K-matrix give the stationary wave solution of the Schr\"{o}dinger equation for
one-dimensional scattering problems. The K-matrix is defined as a Hermite matrix and its poles appear
on the real axis of the complex energy.
The phase shift can be defined using the S-matrix. When analyzing resonances using the phase shift
$\delta$, it is considered that the presence of a resonance significantly increases the phase shift
$\delta$ near the resonance energy and that the phase shift crosses $\delta=\frac{\pi}{2}$ at the
resonance energy.
The poles of the K-matrix are equal to the energy at which $\delta=\frac{\pi}{2}$.
The K-matrix can be represented using the S-matrix, but the poles of each can be related to each other
or independent~\cite{jost-class}. Therefore, there are still many unclear aspects of resonance in terms of its
definition and classification. 

The Jost function is a function of complex energy given as the coefficient which connects the regular
and irregular solutions in the fundamental differential equation of a quantum system based on the
Hamiltonian of the system, such as the Schr\"{o}dinger equation.
It is known that the Jost function can be used to define the S-matrix, and that the zeros of the Jost
function on the complex energy plane provide the complex eigenvalues of the system as the poles of
the S-matrix.
The Jost function was originally formulated for the single-channel system~\cite{jost},
but it can also be defined for multi-channel systems~\cite{Rakityansky, jost-hfb, JostRPA}, since the Jost function can be given as a coefficient function which connects
the regular and irregular solutions in systems where the fundamental equation is given by the
differential equation. In Ref.\cite{jost-hfb} we extended the Jost function within the framework of the Hartree-Fock-
Bogoliubov (HFB) method. By focusing on S- and K-matrix poles in nucleon scattering targeting open-shell nuclei,
we have attempted to analyze and classify the resonances.
As a result, it is found that in nucleon scattering within the framework of HFB theory, there exist
``shape resonances'' formed by centrifugal force potentials and two types of ``quasiparticle resonances''
caused by pair-correlation effects. In such ``resonance states'', the S- and the K-matrix poles are
found to exist simultaneously.
On the other hand, the S- and K-matrix poles can exist independently, in which case the behavior
of the scattering wave function is not characteristic of a resonant state, especially inside the nucleus.
In Ref. \cite{JostRPA}, the Jost function was further extended within the framework of Random-Phase-Approximation (RPA)
theory (hereafter this extended method will be referred to as the Jost-RPA method), and applied to the calculation of the $E1$ excitation modes of $^{16}$O.
As a result, the poles corresponding to collective excitation modes such as the giant resonance were
successfully found on the complex energy plane. 

The S-matrix can be defined by using the Jost function due to the fact that the wave function obtained by
multiplying the regular solution of the system by the inverse of the Jost function satisfies the scattering
boundary conditions.
In Ref.\cite{JostRPA}, it was shown that the poles of the inverse matrix of the Jost function
(the zero point of the determinant of the Jost function) in the representation of the Green function which
gives the RPA response function correspond to the peak of the strength function of the $E1$ excitation mode,
but the representation of the S-matrix using the Jost function has not been explicitly shown.
If we focus only on the asymptotic of the wave function and the definition of the Jost function,
it is most natural to consider that the S-matrix is defined as the Jost function
($\bvec{\mathcal{J}}^{(-)}$) as the coefficient of the outgoing wave multiplied by the inverse of
the Jost function as the coefficient of the incoming wave ($\bvec{\mathcal{J}}^{(+)}$) as
$\bvec{\mathcal{J}}^{(+)-1}\bvec{\mathcal{J}}^{(-)}$. However, the S-matrix defined in this way does
not satisfy unitarity. It is essential to define the S-matrix which satisfies unitarity in order
to analyze and discuss the classification of resonances using S- and K-matrices.
The S-matrix which satisfies unitarity can be diagonalised to define eigenphase shifts.
It may be possible to examine the detailed properties of a physical quantity (such as RPA strength
functions, transition densities, wave functions, etc.) by decomposing it into ``states'' defined
by the eigenphase shift~\cite{Rakityansky,Rodberg,Carew,Lee}.

In this paper, we first present how to define and derive the S-matrix which satisfies unitarity
in the Jost-RPA method (Sec.\ref{Smatder}). 
Then, we show how to decompose the RPA strength function by the states defined for each
eigenphase shift obtained by diagonalisation the S-matrix in Sec.\ref{eigenphasedecom}. 
However, the eigenphase shift does not necessarily correspond to the RPA eigenstates,
since the original S-matrix is made to give boundary conditions that connect to free
particle states at $r\to\infty$ (scattering boundary conditions). Therefore,
in Sec.\ref{S1eigenphase}, we propose how to give the eigenphase shifts corresponding
to the RPA eigenstates. 
In Sec.\ref{isivstr}, the expression of the formulae for the explicit treatment of isospin
dependence is explained. In Sec.\ref{noclmstr}, the S-matrix poles and eigenphase shift
transition densities for isoscalar and isovector modes (which are obtained independently
when the Coulomb interaction is neglected) are described. 
In Sec.\ref{res-dis}, we present numerical results of the application of the method developed
in this paper to the quadrupole excitation $^{16}$O.

\section{Theory}

\subsection{Derivation of the S-matrix in the Jost-RPA}
\label{Smatder}

As shown in Ref.\cite{JostRPA}, the Jost function $\bvec{\mathcal{J}}^{(\pm)}$ in
the extended Jost function method within RPA theory (Jost-RPA method) is defined as
the coefficient function which connects $\bvec{\Phi}^{(r)}$ and $\bvec{\Phi}^{(\pm)}$ as
\begin{eqnarray}
  \bvec{\Phi}^{(r)\mathsf{T}}
  =
  \frac{1}{2}
  \left(
  \bvec{\mathcal{J}}^{(+)}
  \bvec{\Phi}^{(-)\mathsf{T}}
  +
  \bvec{\mathcal{J}}^{(-)}
  \bvec{\Phi}^{(+)\mathsf{T}}
  \right)
\end{eqnarray}
where $\bvec{\Phi}^{(r)}$ and $\bvec{\Phi}^{(\pm)}$ are the regular and irregular solutions of
the coordinate space representation RPA simultaneous differential equations, respectively. 
When the configuration number of the particle-hole excitation in the RPA is given by $N=N_n+N_p$,
where $N_n$ and $N_p$ are the configuration numbers for neutrons and protons, respectively,
$\bvec{\Phi}^{(r)}$, $\bvec{\Phi}^{(\pm)}$ and $\bvec{\mathcal{J}}^{(\pm)}$ are given in a
$2N\times2N$ matrix form. Factor $2$ is due to the fact that the positive and negative energy
solutions exist correlated with each other. 
For example, a particle-hole excitation configuration of $^{16}$O quadrupole excitation is
shown in Table \ref{table1}. In this case, $N_n = 8$ and $N_p = 8$ for neutrons and protons
respectively, which means $N = N_n + N_p = 16$. 

From the asymptotic in the limit of $r\to\infty$ satisfied by the irregular solutions,
it seems that $\bvec{\mathcal{J}}^{(+)-1}\bvec{\mathcal{J}}^{(-)}$ can be defined as an
S-matrix if the scattering wave function $\bvec{\Psi}^{(+)}$ is defined as.
\begin{eqnarray}
  \bvec{\Psi}^{(+)\mathsf{T}}
  \equiv
  \frac{1}{2}
  \left[
  \bvec{\Phi}^{(-)\mathsf{T}}
  +
  \left(
  \bvec{\mathcal{J}}^{(+)-1}
  \bvec{\mathcal{J}}^{(-)}
  \right)
  \bvec{\Phi}^{(+)\mathsf{T}}
  \right],
\end{eqnarray}
however, $\bvec{\mathcal{J}}^{(+)-1}\bvec{\mathcal{J}}^{(-)}$ does not satisfy
the unitarity (which the S-matrix must satisfy in the scattering problem). 

To solve this problem, we first focus on the process of deriving the RPA
Green's function in the Jost-RPA method. 

The RPA Green's function in the Jost-RPA framework is represented by
\begin{eqnarray}
  &&
  \bvec{\mathcal{G}}^{(\pm)}(r,r')
  \nonumber\\
  &&=
  \mp i\frac{2m}{\hbar^2}
  \left[
    \theta(r-r')
    \bvec{\Phi}^{(\pm)}(r)
    \bvec{\mathcal{K}}
    \left(
    \bvec{\mathcal{J}}^{(\pm)-1}
    \right)
    \bvec{\Phi}^{(r)\mathsf{T}}(r')
    \right.
    \nonumber\\
    &&
    \left.
    +
    \theta(r'-r)
    \bvec{\Phi}^{(r)}(r)
    \left(
    \bvec{\mathcal{J}}^{(\pm)-1}
    \right)^{\mathsf{T}}
    \bvec{\mathcal{K}}
    \bvec{\Phi}^{(\pm)\mathsf{T}}(r')
    \right]
  \label{Greendef1}
\end{eqnarray}
$\bvec{\mathcal{K}}$ is a diagonal matrix with the momentum component, where the momentum
component is expressed as $k_{1,\alpha}=\sqrt{\frac{2m}{\hbar^2}(e_\alpha+E)}$ or 
$k_{2,\alpha}=\sqrt{\frac{2m}{\hbar^2}(e_\alpha-E)}$ corresponding to the positive or negative
energy solution, respectively, where $e_\alpha$ is the hole state energy and $E$ is the
excitation energy of the system, $\alpha$ is the subscript which represents the p-h
excitation configuration shown in Table \ref{table1}.

The conditions for the Green's function to be given in the form Eq.(\ref{Greendef1})
require that equations (A.6) and (A.7) given in the Appendix of \cite{JostRPA} are satisfied.
Therefore, the Jost function satisfies the following equation,
\begin{eqnarray}
  &&
  \pm\frac{1}{i}
  \left[
    \bvec{\Phi}^{(\pm)}(r)
    \bvec{\mathcal{K}}
    \left(
    \bvec{\mathcal{J}}^{(\pm)-1}
    \right)
    \del{\bvec{\Phi}^{(r)\mathsf{T}}(r)}{r}
    \right.
    \nonumber\\
    &&
    \left.
    -
    \bvec{\Phi}^{(r)}(r)
    \left(
    \bvec{\mathcal{J}}^{(\pm)-1}
    \right)^{\mathsf{T}}
    \bvec{\mathcal{K}}
    \del{\bvec{\Phi}^{(\pm)\mathsf{T}}(r)}{r}
    \right]
  =
  -\bvec{1}.
\end{eqnarray}
and
\begin{eqnarray}
  &&
  \left[
    \bvec{\Phi}^{(\pm)}(r)
    \bvec{\mathcal{K}}
    \left(
    \bvec{\mathcal{J}}^{(\pm)-1}
    \right)
    \bvec{\Phi}^{(r)\mathsf{T}}(r)
    \right.
    \nonumber\\
    &&
    \left.
    -
    \bvec{\Phi}^{(r)}(r)
    \left(
    \bvec{\mathcal{J}}^{(\pm)-1}
    \right)^{\mathsf{T}}
    \bvec{\mathcal{K}}
    \bvec{\Phi}^{(\pm)\mathsf{T}}(r)
    \right]
  =
  \bvec{0}.
\end{eqnarray}
Substituting the relationship between regular and irregular solutions (which is the definition
of the Jost function) into these equations and considering the limit of $r\to \infty$,
one can obtain the following equation
\begin{eqnarray}
  &&
  \bvec{\mathcal{K}}^{\frac{1}{2}}
  \left(
  \bvec{\mathcal{J}}^{(\pm)-1}
  \right)
  \left(
  \bvec{\mathcal{J}}^{(\mp)}
  \right)
  \bvec{\mathcal{K}}^{-\frac{1}{2}}
  \nonumber\\
  &&
  =
  \left[
    \bvec{\mathcal{K}}^{\frac{1}{2}}
    \left(
    \bvec{\mathcal{J}}^{(\pm)-1}
    \right)
    \left(
    \bvec{\mathcal{J}}^{(\mp)}
    \right)
    \bvec{\mathcal{K}}^{-\frac{1}{2}}
    \right]^{\mathsf{T}}.
\end{eqnarray}
This formula shows that the left-hand side is a symmetric matrix, which we define as the
``S-matrix'' (although this is not the S-matrix which satisfies the unitarity used in the
scattering problem) as
\begin{eqnarray}
  \bvec{\mathcal{S}}(\bvec{\mathcal{K}})
  &\equiv&
  \bvec{\mathcal{K}}^{\frac{1}{2}}
  \bvec{\mathcal{J}}^{(+)-1}
  \bvec{\mathcal{J}}^{(-)}
  \bvec{\mathcal{K}}^{-\frac{1}{2}}
  =
  \bvec{\mathcal{S}}^{\mathsf{T}}(\bvec{\mathcal{K}}).
  \label{Smat-def}
\end{eqnarray}
Note that the S-matrix defined by multiplying $\bvec{\mathcal{K}}^{\frac{1}{2}}$
and $\bvec{\mathcal{K}}^{-\frac{1}{2}}$ from both sides in this way is sometimes
called the ``flux adjusted S-matrix''~\cite{Rakityansky}, and an ``adjust factor'' for
mass must also be applied to define the S-matrix if the mass of the particle
changes depending on the channel. 

Using this $\bvec{\mathcal{S}}$ with the symmetric property
and the regular and irregular solutions redefined as
\begin{eqnarray}
  \widehat{\bvec{\Phi}}^{(r)\mathsf{T}}(\bvec{\mathcal{K}})
  &=&
  \bvec{\mathcal{K}}^{\frac{1}{2}}
  \bvec{\Phi}^{(\pm)\mathsf{T}}(\bvec{\mathcal{K}})
  \nonumber\\
  &=&
  \frac{1}{2}
  \left[
    \bvec{\mathcal{J}}^{(+)}
    \widehat{\bvec{\Phi}}^{(-)\mathsf{T}}(\bvec{\mathcal{K}})
    +
    \bvec{\mathcal{J}}^{(-)}
    \widehat{\bvec{\Phi}}^{(+)\mathsf{T}}(\bvec{\mathcal{K}})
    \right]
  \\
  \widehat{\bvec{\Phi}}^{(\pm)\mathsf{T}}(\bvec{\mathcal{K}})
  &=&
  \bvec{\mathcal{K}}^{\frac{1}{2}}
  \bvec{\Phi}^{(\pm)\mathsf{T}}(\bvec{\mathcal{K}}),  
\end{eqnarray}
the Green's function Eq.(\ref{Greendef1}) can be expressed as
\begin{eqnarray}
  \bvec{\mathcal{G}}^{(+)}(r,r')
  &&=
  -i\frac{m}{\hbar^2}
  \left[
    \theta(r-r')
    \widehat{\bvec{\Phi}}^{(+)}(r)
    \widehat{\bvec{\Phi}}^{(-)\mathsf{T}}(r')
    \right.
    \nonumber\\
    &&
    \left.
    +
    \theta(r'-r)
    \widehat{\bvec{\Phi}}^{(-)}(r)
    \widehat{\bvec{\Phi}}^{(+)\mathsf{T}}(r')
    \right.
    \nonumber\\
    &&
    \left.
    +
    \theta(r'-r)
    \widehat{\bvec{\Phi}}^{(+)}(r)
    \bvec{\mathcal{S}}
    \widehat{\bvec{\Phi}}^{(+)\mathsf{T}}(r')
    \right]
  \label{Greendef2}
\end{eqnarray}
Since ``the unitarity which the S-matrix must satisfy'' and the
``analytic continuation between Riemann sheets of the complex energy plane''
are closely related, we next show the properties of the complex
conjugate of the Green's function. 
In RPA theory, it is considered to have analytic continuation with various
Riemann sheets where resonances exist at branch-cut lines above various
thresholds in the positive and negative energy regions respectively.
In order to express explicitly the contributions and correlations relating
to the positive and negative energy solutions in the expression of the
Green's function, we make the following equation transformations.

When $\bvec{\mathcal{K}}_1$ and $\bvec{\mathcal{K}}_2$ are $N\times N$ diagonal matrices
with $k_{1,\alpha}$ and $k_{2,\alpha}$ as matrix components, respectively, and $\bvec{\mathcal{K}}$
is represented as
\begin{eqnarray}
  \bvec{\mathcal{K}}
  =
  \begin{pmatrix}
    \bvec{\mathcal{K}}_1 & \bvec{0} \\
    \bvec{0} & \bvec{\mathcal{K}}_2
  \end{pmatrix},
\end{eqnarray}
we represent $\widehat{\bvec{\Phi}}^{(\pm)}$ as
\begin{eqnarray}
  \widehat{\bvec{\Phi}}^{(\pm)}
  &=&
  \begin{pmatrix}
    \widehat{\bvec{\phi}}^{(\pm 1)} &
    \widehat{\bvec{\phi}}^{(\pm 2)}
  \end{pmatrix}
  \label{phipm}
\end{eqnarray}
where $\widehat{\bvec{\phi}}^{(\pm 1)}$ and $\widehat{\bvec{\phi}}^{(\pm 2)}$
are $2N\times N$ matrices.
The Jost function can also be expressed as
\begin{eqnarray}
  \bvec{\mathcal{J}}^{(\pm)}
  =
  \begin{pmatrix}
    \bvec{\mathcal{J}}_{11}^{(\pm)} & \bvec{\mathcal{J}}_{12}^{(\pm)} \\
    \bvec{\mathcal{J}}_{21}^{(\pm)} & \bvec{\mathcal{J}}_{22}^{(\pm)} 
  \end{pmatrix},
\end{eqnarray}
and the ``S-matrix'' is expressed as
\begin{eqnarray}
  \bvec{\mathcal{S}}
  =
  \begin{pmatrix}
    \bvec{\mathcal{S}}_{11} & \bvec{\mathcal{S}}_{12} \\
    \bvec{\mathcal{S}}_{21} & \bvec{\mathcal{S}}_{22} 
  \end{pmatrix}
\end{eqnarray}
with
\begin{eqnarray}
  \bvec{\mathcal{S}}_{11}
  &=&
  \bvec{\mathcal{K}}_1^{\frac{1}{2}}
  \left(\bvec{\mathcal{J}}^{(+)}_{11}
  -
  \bvec{\mathcal{J}}^{(+)}_{12}
  \bvec{\mathcal{J}}^{(+)-1}_{22}
  \bvec{\mathcal{J}}^{(+)}_{21}\right)^{-1}
  \nonumber\\
  &&
  \left(\bvec{\mathcal{J}}^{(-)}_{11}
  -
  \bvec{\mathcal{J}}^{(+)}_{12}
  \bvec{\mathcal{J}}^{(+)-1}_{22}
  \bvec{\mathcal{J}}^{(-)}_{21}\right)
  \bvec{\mathcal{K}}_1^{-\frac{1}{2}}
  \label{S11def}
  \\
  \bvec{\mathcal{S}}_{12}
  &=&
  \bvec{\mathcal{K}}_1^{\frac{1}{2}}
  \left(\bvec{\mathcal{J}}^{(+)}_{11}
  -
  \bvec{\mathcal{J}}^{(+)}_{12}
  \bvec{\mathcal{J}}^{(+)-1}_{22}
  \bvec{\mathcal{J}}^{(+)}_{21}\right)^{-1}
  \nonumber\\
  &&
  \left(\bvec{\mathcal{J}}^{(-)}_{12}
  -
  \bvec{\mathcal{J}}^{(+)}_{12}
  \bvec{\mathcal{J}}^{(+)-1}_{22}
  \bvec{\mathcal{J}}^{(-)}_{22}\right)
  \bvec{\mathcal{K}}_2^{-\frac{1}{2}}
  \\
  \bvec{\mathcal{S}}_{21}
  &=&
  \bvec{\mathcal{K}}_2^{\frac{1}{2}}
  \bvec{\mathcal{J}}^{(+)-1}_{22}
  \nonumber\\
  &&
  \left(
    \bvec{\mathcal{J}}^{(-)}_{21}
    -
    \bvec{\mathcal{J}}^{(+)}_{21}
    \bvec{\mathcal{K}}_1^{-\frac{1}{2}}
    \bvec{\mathcal{S}}_{11}
    \bvec{\mathcal{K}}_1^{\frac{1}{2}}
    \right)
  \bvec{\mathcal{K}}_1^{-\frac{1}{2}}
  \\
  \bvec{\mathcal{S}}_{22}
  &=&
  \bvec{\mathcal{K}}_2^{\frac{1}{2}}
  \bvec{\mathcal{J}}^{(+)-1}_{22}
  \nonumber\\
  &&
  \left(
    \bvec{\mathcal{J}}^{(-)}_{22}
    -
    \bvec{\mathcal{J}}^{(+)}_{21}
    \bvec{\mathcal{K}}_1^{-\frac{1}{2}}
    \bvec{\mathcal{S}}_{12}
    \bvec{\mathcal{K}}_2^{\frac{1}{2}}
    \right)
  \bvec{\mathcal{K}}_2^{-\frac{1}{2}}
\end{eqnarray}
The $\det \bvec{\mathcal{J}}^{(+)}$ which gives the poles of
the S-matrix can be expressed as
\begin{eqnarray}
  &&
  \det \bvec{\mathcal{J}}^{(+)}
  \nonumber
  \\
  &&
  =
  \det
  \bvec{\mathcal{J}}^{(+)}_{22}
  \det
  \left(\bvec{\mathcal{J}}^{(+)}_{11}
  -
  \bvec{\mathcal{J}}^{(+)}_{12}
  \bvec{\mathcal{J}}^{(+)-1}_{22}
  \bvec{\mathcal{J}}^{(+)}_{21}\right)
  \label{detJ1}
  \\
  &&
  =
  \det
  \bvec{\mathcal{J}}^{(+)}_{11}
  \det
  \left(\bvec{\mathcal{J}}^{(+)}_{22}
  -
  \bvec{\mathcal{J}}^{(+)}_{21}
  \bvec{\mathcal{J}}^{(+)-1}_{11}
  \bvec{\mathcal{J}}^{(+)}_{12}\right).
  \label{detJ2}
\end{eqnarray}
In particular, the complex energy $E$ satisfying
\begin{eqnarray}
  \det
  \left(\bvec{\mathcal{J}}^{(+)}_{11}
  -
  \bvec{\mathcal{J}}^{(+)}_{12}
  \bvec{\mathcal{J}}^{(+)-1}_{22}
  \bvec{\mathcal{J}}^{(+)}_{21}
  \right)
  =0
  \label{detJ11}
\end{eqnarray}
gives the S-matrix pole in the positive energy region.

By using these representation, the Green's function is represented as
\begin{eqnarray}
  \bvec{\mathcal{G}}^{(+)}(r,r')
  &&=
  -i\frac{m}{\hbar^2}
  \left[
    \theta(r-r')
    \widehat{\bvec{\phi}}^{(+1)}(r)
    \widehat{\bvec{\phi}}^{(-1)\mathsf{T}}(r')
    \right.
    \nonumber\\
    &&
    \left.
    +
    \theta(r'-r)
    \widehat{\bvec{\phi}}^{(-1)}(r)
    \widehat{\bvec{\phi}}^{(+1)\mathsf{T}}(r')
    \right.
    \nonumber\\
    &&
    +
    \left.
    \theta(r-r')
    \widehat{\bvec{\phi}}^{(+2)}(r)
    \widehat{\bvec{\phi}}^{(-2)\mathsf{T}}(r')
    \right.
    \nonumber\\
    &&
    \left.
    +
    \theta(r'-r)
    \widehat{\bvec{\phi}}^{(-2)}(r)
    \widehat{\bvec{\phi}}^{(+2)\mathsf{T}}(r')
    \right.
    \nonumber\\
    &&
    \left.
    +
    \widehat{\bvec{\phi}}^{(+1)}(r)
    \bvec{\mathcal{S}}_{11}
    \widehat{\bvec{\phi}}^{(+1)\mathsf{T}}(r')
    \right.
    \nonumber\\
    &&
    \left.
    +
    \widehat{\bvec{\phi}}^{(+1)}(r)
    \bvec{\mathcal{S}}_{12}
    \widehat{\bvec{\phi}}^{(+2)\mathsf{T}}(r')
    \right.
    \nonumber\\
    &&
    \left.
    +
    \widehat{\bvec{\phi}}^{(+2)}(r)
    \bvec{\mathcal{S}}_{21}
    \widehat{\bvec{\phi}}^{(+1)\mathsf{T}}(r')
    \right.
    \nonumber\\
    &&
    \left.
    +
    \widehat{\bvec{\phi}}^{(+2)}(r)
    \bvec{\mathcal{S}}_{22}
    \widehat{\bvec{\phi}}^{(+2)\mathsf{T}}(r')
    \right]
  \label{Greendef3}
\end{eqnarray}
In the RPA theory, solutions are symmetrically given between the positive and negative
energy regions, so the analytic continuation of the Riemann sheets of the complex energy
plane is also symmetric between the positive and negative energy regions.
Therefore, the properties of complex conjugation related to the analytic continuation
in the positive energy region are presented below.

In the positive energy region (Re $E>0$), when Re $E \leqq -e_{\alpha_c}$
($\alpha_c$ is a certain configuration), the momentum have the following properties,
\begin{eqnarray}
  k_{1,\alpha}(E^*)
  &=&
  \left\{
  \begin{array}{cc}
    k_{1,\alpha}^*(E) & \mbox{for $\alpha\leqq\alpha_c$} \\
    -k_{1,\alpha}^*(E) & \mbox{for $\alpha > \alpha_c$}
  \end{array}
  \right.
  \label{k1prop}
  \\
  k_{2,\alpha}(E^*)
  &=&
  -k_{2,\alpha}^*(E) \hspace{10pt}\mbox{for all $\alpha(\in 1,2,\cdots N)$}.
  \label{k2prop}
\end{eqnarray}
Although trivial, these properties of momentum indicate that on the real
axis of complex energy $E$, $k_{1,\alpha}$ for $\alpha > \alpha_c$ and
$k_{2,\alpha}$ are pure imaginary numbers and $k_{1,\alpha}$ for
$\alpha\leqq\alpha_c$ is a real number.

To express the nature of the complex conjugation of $\widehat{\bvec{\Phi}}^{(\pm)}$
as a function of complex energy $E$, the column components of $\widehat{\bvec{\phi}}^{(\pm 1)}$ in Eq.(\ref{phipm})
are delimited by $\alpha_c$ as
\begin{eqnarray}
  \widehat{\bvec{\phi}}^{(\pm 1)}
  &=&
  \begin{pmatrix}
    \widehat{\bvec{\phi}}^{(\pm 1c)} &
    \widehat{\bvec{\phi}}^{(\pm 1d)} 
  \end{pmatrix}
  \label{phipm2}
\end{eqnarray}
where $\widehat{\bvec{\phi}}^{(\pm 1c)}$ and $\widehat{\bvec{\phi}}^{(\pm 1d)}$ are
$2N\times\alpha_c$ and $2N\times(N-\alpha_c)$ matrices, respectively.

Due to the properties of momentum given by Eqs.(\ref{k1prop}) and (\ref{k2prop}),
the complex conjugate of $\widehat{\bvec{\Phi}}^{(\pm)}$ as a function of complex
energy $E$ is given by
\begin{eqnarray}
  \widehat{\bvec{\Phi}}^{(\pm)*}(E^*)
  &=&
  \begin{pmatrix}
    \widehat{\bvec{\phi}}^{(\pm 1)*}(E^*) &
    \widehat{\bvec{\phi}}^{(\pm 2)*}(E^*)
  \end{pmatrix}
  \label{phipm2-conjg}
\end{eqnarray}
with
\begin{eqnarray}
  \widehat{\bvec{\phi}}^{(\pm 1c)*}(E^*)
  &=&
  \widehat{\bvec{\phi}}^{(\mp 1c)}(E)
  \\
  \widehat{\bvec{\phi}}^{(\pm 1d)*}(E^*)
  &=&
  \widehat{\bvec{\phi}}^{(\pm 1d)}(E)
  i\bvec{\eta}_d
  \\
  \widehat{\bvec{\phi}}^{(\pm 2)*}(E^*)
  &=&
  \widehat{\bvec{\phi}}^{(\pm 2)}(E)
  i\bvec{\eta}
\end{eqnarray}
where $\bvec{\eta}$ is a $N\times N$ diagonal matrix with $(-)^{l^{(\alpha)}}$ as a matrix element
($l^{(\alpha)}$ is the particle angular momentum number of a particle-hole excitation
configuration component $\alpha$), and $\bvec{\eta}_d$ is the same, but 
$(N-\alpha_c)\times(N-\alpha_c)$ diagonal matrix for $\alpha>\alpha_c$.

Also the complex conjugate of the regular solution $\widehat{\bvec{\Phi}}^{(r)}(E)$ is given by
\begin{eqnarray}
  \widehat{\bvec{\Phi}}^{(r)*}(E^*)
  &=&
  \begin{pmatrix}
    \widehat{\bvec{\phi}}^{(r 1)*}(E^*) &
    \widehat{\bvec{\phi}}^{(r 2)*}(E^*)
  \end{pmatrix}
  \label{phir2-conjg}
\end{eqnarray}
with
\begin{eqnarray}
  \widehat{\bvec{\phi}}^{(r 1)}
  &=&
  \begin{pmatrix}
    \widehat{\bvec{\phi}}^{(r 1c)} &
    \widehat{\bvec{\phi}}^{(r 1d)} 
  \end{pmatrix}
  \label{phir2}
\end{eqnarray}
and
\begin{eqnarray}
  \widehat{\bvec{\phi}}^{(r 1c)*}(E^*)
  &=&
  \widehat{\bvec{\phi}}^{(r 1c)}(E)
  \\
  \widehat{\bvec{\phi}}^{(r 1d)*}(E^*)
  &=&
  \widehat{\bvec{\phi}}^{(r 1d)}(E)
  i\bvec{\eta}_d
  \\
  \widehat{\bvec{\phi}}^{(r 2)*}(E^*)
  &=&
  \widehat{\bvec{\phi}}^{(r 2)}(E)
  i\bvec{\eta},
\end{eqnarray}
By applying these properties of solutions to the definition of the Jost function,
we can obtain the complex conjugation properties of the ``S-matrix''.
To show the complex conjugate nature of the ``S-matrix'',
$\bvec{\mathcal{S}}_{11}$, $\bvec{\mathcal{S}}_{12}$ and $\bvec{\mathcal{S}}_{21}$
are divided into block matrices bounded by $\alpha=\alpha_c$ as follows.
\begin{eqnarray}
  \bvec{\mathcal{S}}_{11}
  &=&
  \begin{pmatrix}
    \bvec{\mathcal{S}}_{11}^{cc} & \bvec{\mathcal{S}}_{11}^{cd} \\
    \bvec{\mathcal{S}}_{11}^{dc} & \bvec{\mathcal{S}}_{11}^{dd}
  \end{pmatrix},
  \hspace{10pt}
  \bvec{\mathcal{S}}_{12}
  =
  \begin{pmatrix}
    \bvec{\mathcal{S}}_{12}^{cd}  \\
    \bvec{\mathcal{S}}_{12}^{dd} 
  \end{pmatrix}
  \\
  \bvec{\mathcal{S}}_{21}
  &=&
  \begin{pmatrix}
    \bvec{\mathcal{S}}_{21}^{dc} & \bvec{\mathcal{S}}_{21}^{dd} 
  \end{pmatrix}
\end{eqnarray}
where $\bvec{\mathcal{S}}_{11}^{cc}$ is the $\alpha_c\times\alpha_c$ matrix,
$\bvec{\mathcal{S}}_{11}^{cd}$ is the $\alpha_c\times(N-\alpha_c)$ matrix,
$\bvec{\mathcal{S}}_{11}^{dc}$ is the $(N-\alpha_c)\times\alpha_c$ matrix,
$\bvec{\mathcal{S}}_{11}^{dd}$ is the $(N-\alpha_c)\times(N-\alpha_c)$ matrix,
$\bvec{\mathcal{S}}_{12}^{cd}$ is the $\alpha_c\times N$ matrix,
$\bvec{\mathcal{S}}_{12}^{dd}$ is the $(N-\alpha_c)\times N$ matrix,
$\bvec{\mathcal{S}}_{21}^{dc}$ is the $N\times\alpha_c$ matrix and 
$\bvec{\mathcal{S}}_{21}^{dd}$ is the $N\times(N-\alpha_c)$ matrix.

To distinguish between above-threshold terms, below-threshold terms and their correlated terms in the ``S-matrix'',
The block matrices are then further rearranged as 
\begin{eqnarray}
  &&\bvec{\mathcal{S}}
  =
  \begin{pmatrix}
    \bvec{\mathcal{S}}_{cc} & \bvec{\mathcal{S}}_{cd} \\
    \bvec{\mathcal{S}}_{dc} & \bvec{\mathcal{S}}_{dd} 
  \end{pmatrix}
  \\
  &&
  \left\{
  \begin{array}{ll}
    \bvec{\mathcal{S}}_{cc}
    \equiv
    \bvec{\mathcal{S}}_{11}^{cc}
    &
    \mbox{($\alpha_c\times\alpha_c$ matrix)}
    \\
    \bvec{\mathcal{S}}_{cd}
    \equiv
    \begin{pmatrix}
      \bvec{\mathcal{S}}_{11}^{cd} & \bvec{\mathcal{S}}_{12}^{cd}
    \end{pmatrix}
    &
    \mbox{($\alpha_c\times(2N-\alpha_c)$ matrix)}
    \\
    \bvec{\mathcal{S}}_{dc}
    \equiv
    \begin{pmatrix}
      \bvec{\mathcal{S}}_{11}^{dc} \\
      \bvec{\mathcal{S}}_{21}^{dc}
    \end{pmatrix}
    &
    \mbox{($(2N-\alpha_c)\times\alpha_c$ matrix)}
    \\
    \bvec{\mathcal{S}}_{dd}
    \equiv
    \begin{pmatrix}
      \bvec{\mathcal{S}}_{11}^{dd} & \bvec{\mathcal{S}}_{12}^{dd} \\
      \bvec{\mathcal{S}}_{21}^{dd} & \bvec{\mathcal{S}}_{22}^{dd}
    \end{pmatrix}
    &
    \mbox{($(2N-\alpha_c)\times(2N-\alpha_c)$ matrix)}
  \end{array}
  \right.
  \label{Sccdef}
\end{eqnarray}
The complex conjugates of these rearranged block matrices
are given as
\begin{eqnarray}
  \bvec{\mathcal{S}}^*_{cc}(E^*)
  &=&
  \bvec{\mathcal{S}}^{-1}_{cc}(E)
  \label{Scc-conjg}
  \\
  \bvec{\mathcal{S}}^*_{cd}(E^*)
  &=&
  -
  \bvec{\mathcal{S}}^{-1}_{cc}(E)
  \bvec{\mathcal{S}}_{cd}(E)
  i\widetilde{\bvec{\eta}}
  \\
  \bvec{\mathcal{S}}^*_{dc}(E^*)
  &=&
  -
  i\widetilde{\bvec{\eta}}
  \bvec{\mathcal{S}}_{dc}(E)
  \bvec{\mathcal{S}}^{-1}_{cc}(E)
  \\
  \bvec{\mathcal{S}}^*_{dd}(E^*)
  &=&
  \widetilde{\bvec{\eta}}
  \bvec{\mathcal{S}}_{dd}(E)
  \widetilde{\bvec{\eta}}
  \nonumber\\
  &&
  -
  \widetilde{\bvec{\eta}}
  \bvec{\mathcal{S}}_{dc}(E)
  \bvec{\mathcal{S}}^{-1}_{cc}(E)
  \bvec{\mathcal{S}}_{cd}(E)
  \widetilde{\bvec{\eta}}
\end{eqnarray}
respectively, where $\widetilde{\bvec{\eta}}$ is the $(2N-\alpha_c)\times(2N-\alpha_c)$ diagonal
matrix which is defined by
\begin{eqnarray}
  \widetilde{\bvec{\eta}}
  =
  \begin{pmatrix}
    \bvec{\eta}_d & \bvec{0} \\
    \bvec{0} & \bvec{\eta}
  \end{pmatrix}.
\end{eqnarray}
If the irregular solution is also rearranged as
\begin{eqnarray}
  \widehat{\bvec{\Phi}}^{(\pm)}
  &=&
  \begin{pmatrix}
    \widehat{\bvec{\phi}}^{(\pm c)} &
    \widehat{\bvec{\phi}}^{(\pm d)}
  \end{pmatrix}
  \label{phipm3}
\end{eqnarray}
with
\begin{eqnarray}
  \widehat{\bvec{\phi}}^{(\pm c)}
  &\equiv&
  \widehat{\bvec{\phi}}^{(\pm 1c)}
  \\
  \widehat{\bvec{\phi}}^{(\pm d)}
  &\equiv&
  \begin{pmatrix}
    \widehat{\bvec{\phi}}^{(\pm 1d)} &
    \widehat{\bvec{\phi}}^{(\pm 2)}
  \end{pmatrix}
\end{eqnarray}
(note that $\widehat{\bvec{\phi}}^{(\pm c)}$ and $\widehat{\bvec{\phi}}^{(\pm d)}$ are
$2N\times\alpha_c$ and $2N\times(2N-\alpha_c)$ matrices, respectively), 
the Green's function Eq.(\ref{Greendef3}) can be expressed as
\begin{eqnarray}
  \bvec{\mathcal{G}}^{(+)}(r,r')
  &&=
  -i\frac{m}{\hbar^2}
  \left[
    \theta(r-r')
    \widehat{\bvec{\phi}}^{(+c)}(r)
    \widehat{\bvec{\phi}}^{(-c)\mathsf{T}}(r')
    \right.
    \nonumber\\
    &&
    \left.
    +
    \theta(r'-r)
    \widehat{\bvec{\phi}}^{(-c)}(r)
    \widehat{\bvec{\phi}}^{(+c)\mathsf{T}}(r')
    \right.
    \nonumber\\
    &&
    \left.
    +
    \theta(r-r')
    \widehat{\bvec{\phi}}^{(+d)}(r)
    \widehat{\bvec{\phi}}^{(-d)\mathsf{T}}(r')
    \right.
    \nonumber\\
    &&
    \left.
    +
    \theta(r'-r)
    \widehat{\bvec{\phi}}^{(-d)}(r)
    \widehat{\bvec{\phi}}^{(+d)\mathsf{T}}(r')
    \right.
    \nonumber\\
    &&
    \left.
    +
    \widehat{\bvec{\phi}}^{(+c)}(r)
    \bvec{\mathcal{S}}_{cc}
    \widehat{\bvec{\phi}}^{(+c)\mathsf{T}}(r')
    \right.
    \nonumber\\
    &&
    \left.
    +
    \widehat{\bvec{\phi}}^{(+c)}(r)
    \bvec{\mathcal{S}}_{cd}
    \widehat{\bvec{\phi}}^{(+d)\mathsf{T}}(r')
    \right.
    \nonumber\\
    &&
    \left.
    +
    \widehat{\bvec{\phi}}^{(+d)}(r)
    \bvec{\mathcal{S}}_{dc}
    \widehat{\bvec{\phi}}^{(+c)\mathsf{T}}(r')
    \right.
    \nonumber\\
    &&
    \left.
    +
    \widehat{\bvec{\phi}}^{(+d)}(r)
    \bvec{\mathcal{S}}_{dd}
    \widehat{\bvec{\phi}}^{(+d)\mathsf{T}}(r')
    \right],
  \label{Greendef4}
\end{eqnarray}
and its complex conjugate is given by
\begin{eqnarray}
  \bvec{\mathcal{G}}^{(+)*}(r,r';E^*)
  &&=
  i\frac{m}{\hbar^2}
  \left[
    \theta(r-r')
    \widehat{\bvec{\phi}}^{(-c)}(r)
    \widehat{\bvec{\phi}}^{(+c)\mathsf{T}}(r')
    \right.
    \nonumber\\
    &&
    \left.
    +
    \theta(r'-r)
    \widehat{\bvec{\phi}}^{(+c)}(r)
    \widehat{\bvec{\phi}}^{(-c)\mathsf{T}}(r')
    \right.
    \nonumber\\
    &&-
    \left.
    \theta(r-r')
    \widehat{\bvec{\phi}}^{(+d)}(r)
    \widehat{\bvec{\phi}}^{(-d)\mathsf{T}}(r')
    \right.
    \nonumber\\
    &&
    \left.
    -
    \theta(r'-r)
    \widehat{\bvec{\phi}}^{(-d)}(r)
    \widehat{\bvec{\phi}}^{(+d)\mathsf{T}}(r')
    \right.
    \nonumber\\
    &&
    \left.
    +
    \widehat{\bvec{\phi}}^{(-c)}(r)
    \bvec{\mathcal{S}}_{cc}^{-1}
    \widehat{\bvec{\phi}}^{(-c)\mathsf{T}}(r')
    \right.
    \nonumber\\
    &&
    \left.
    +
    \widehat{\bvec{\phi}}^{(-c)}(r)
    \bvec{\mathcal{S}}_{cc}^{-1}
    \bvec{\mathcal{S}}_{cd}
    \widehat{\bvec{\phi}}^{(+d)\mathsf{T}}(r')
    \right.
    \nonumber\\
    &&
    \left.
    +
    \widehat{\bvec{\phi}}^{(+d)}(r)
    \bvec{\mathcal{S}}_{dc}
    \bvec{\mathcal{S}}_{cc}^{-1}
    \widehat{\bvec{\phi}}^{(-c)\mathsf{T}}(r')
    \right.
    \nonumber\\
    &&
    \left.
    -
    \widehat{\bvec{\phi}}^{(+d)}(r)
    \bvec{\mathcal{S}}_{dd}
    \widehat{\bvec{\phi}}^{(+d)\mathsf{T}}(r')
    \right.
    \nonumber\\
    &&
    \left.
    +
    \widehat{\bvec{\phi}}^{(+d)}(r)
    \bvec{\mathcal{S}}_{dc}
    \bvec{\mathcal{S}}_{cc}^{-1}
    \bvec{\mathcal{S}}_{cd}
    \widehat{\bvec{\phi}}^{(+d)\mathsf{T}}(r')
    \right].
  \label{Greendef4-conjg}
\end{eqnarray}
Using the properties of the Green's function,
the S-matrix and the complex conjugate of the irregular solution shown above,
one can obtain the ``imaginary'' part of the Green's function as
\begin{eqnarray}
  &&
  \frac{1}{2i}
  \left[
    \bvec{\mathcal{G}}^{(+)}(r,r';E)
    -
    \bvec{\mathcal{G}}^{(+)*}(r,r';E^*)
    \right]
  \nonumber\\
  &&
  =
  -\frac{2m}{\hbar^2}
  \widehat{\bvec{\psi}}^{(+c)}(r;E)
  \widehat{\bvec{\psi}}^{(+c)\dagger}(r';E^*)
  \label{aimag-G}
\end{eqnarray}
with $\widehat{\bvec{\psi}}^{(+c)}(r)$ which is defined as
\begin{eqnarray}
  \widehat{\bvec{\psi}}^{(+c)}(r;E)
  &\equiv&
  \frac{1}{2}
  \left[
    \widehat{\bvec{\phi}}^{(-c)}(r;E)
    +
    \widehat{\bvec{\phi}}^{(+c)}(r;E)
    \bvec{\mathcal{S}}_{cc}(E)
    \right.
    \nonumber\\
    &&
    \left.
    +
    \widehat{\bvec{\phi}}^{(+d)}(r;E)
    \bvec{\mathcal{S}}_{dc}(E)
    \right]
  \label{psic}
\end{eqnarray}
where this wave function is given as the $2N\times\alpha_c$ matrix. 
The complex conjugate of $\widehat{\bvec{\psi}}^{(+c)}(r)$ can be expressed as
\begin{eqnarray}
  \widehat{\bvec{\psi}}^{(+c)\dagger}(r;E^*)
  &=&
  \frac{1}{2}
  \left[
    \widehat{\bvec{\phi}}^{(+c)\mathsf{T}}(r;E)
    +
    \bvec{\mathcal{S}}_{cc}^{-1}(E)
    \widehat{\bvec{\phi}}^{(-c)\mathsf{T}}(r;E)
    \right.
    \nonumber\\
    &&
    \left.
    +
    \bvec{\mathcal{S}}_{cc}^{-1}(E)
    \bvec{\mathcal{S}}_{cd}(E)
    \widehat{\bvec{\phi}}^{(+d)\mathsf{T}}(r;E)
    \right].
\end{eqnarray}
due to the complex conjugation properties of $\bvec{\mathcal{S}}$
and irregular solutions.

Eqs.(\ref{Smat-def}) and (\ref{Scc-conjg}) show that $\bvec{\mathcal{S}}_{cc}$ is a
unitary matrix and satisfies
$\bvec{\mathcal{S}}_{cc}^\dagger \bvec{\mathcal{S}}_{cc}
=\bvec{\mathcal{S}}_{cc}\bvec{\mathcal{S}}_{cc}^\dagger=\bvec{1}$ where $\bvec{1}$ is an
$\alpha_c\times\alpha_c$ unit matrix.

Since $\widehat{\bvec{\psi}}^{(-c)}$ is defined by
$\widehat{\bvec{\psi}}^{(-c)}\equiv\widehat{\bvec{\psi}}^{(+c)*}(E^*)$,
$\widehat{\bvec{\psi}}^{(-c)}$ can be expressed as
\begin{eqnarray}
  \widehat{\bvec{\psi}}^{(-c)\mathsf{T}}(r;E)
  &=&
  \widehat{\bvec{\psi}}^{(+c)\dagger}(r;E^*)
  \\
  &=&
  \frac{1}{2}
  \left[
    \widehat{\bvec{\phi}}^{(+c)\mathsf{T}}(r;E)
    +
    \bvec{\mathcal{S}}_{cc}^{-1}(E)
    \widehat{\bvec{\phi}}^{(-c)\mathsf{T}}(r;E)
    \right.
    \nonumber\\
    &&
    \left.
    +
    \bvec{\mathcal{S}}_{cc}^{-1}(E)
    \bvec{\mathcal{S}}_{cd}(E)
    \widehat{\bvec{\phi}}^{(+d)\mathsf{T}}(r;E)
    \right],
\end{eqnarray}
therefore one can find that $\widehat{\bvec{\psi}}^{(+c)}$ and $\widehat{\bvec{\psi}}^{(-c)}$ are related
as $\widehat{\bvec{\psi}}^{(+c)\mathsf{T}}(E)
=\bvec{\mathcal{S}}_{cc}(E)\widehat{\bvec{\psi}}^{(-c)\mathsf{T}}(E)$. 
This is the same relationship as that between the scattering wavefunction and the S-matrix
in the scattering problem. 

As shown in Ref.\cite{JostRPA}, the RPA response function $R(r,r';E)$ is expressed using a $2N$-dimensional
vector $\vec{\bvec{\varphi}}(r)$ defined by the hole state wave function and a Green's function
$\bvec{\mathcal{G}}^{(+)}(r,r';E)$ as
\begin{eqnarray}
  R(r,r';E)
  &=&
  \vec{\bvec{\varphi}}^{\mathsf{T}}(r)
  \bvec{\mathcal{G}}^{(+)}(r,r';E)
  \vec{\bvec{\varphi}}(r')
  \label{RPAresp}
\end{eqnarray}
Applying Eq. (\ref{aimag-G}) to the definition of the strength function $S_F(E)$ for the external
field $f(r)$ expressed using the RPA response function $R(r,r';E)$, the strength function can be expressed as
\begin{eqnarray}
  &&
  S_F(E)
  \nonumber\\
  &&=
  -\frac{1}{\pi}
  \mbox{ Im }
  \int\int drdr'
  f(r)
  R(r,r';E)
  f(r')
  \nonumber\\
  &&=
  \left[
    \sqrt{
      \frac{1}{\pi}
      \frac{2m}{\hbar^2}
    }
    \int dr
    \vec{\bvec{\varphi}}^{\mathsf{T}}(r)
    f(r)
    \widehat{\bvec{\psi}}^{(+c)}(r;E)
    \right]
  \nonumber\\
  &&\times
  \left[
    \sqrt{
      \frac{1}{\pi}
      \frac{2m}{\hbar^2}
    }
    \int dr'
    \widehat{\bvec{\psi}}^{(+c)\dagger}(r';E^*)
    f(r')
    \vec{\bvec{\varphi}}(r')
    \right]
  \label{RPAstr}
\end{eqnarray}
From the above, we can say that in the framework of RPA theory,
$\widehat{\bvec{\psi}}^{(\pm c)}$ is the scattering wave function representing
the transition to continuum including resonant states, and $\bvec{\mathcal{S}}_{cc}$
is the S-matrix which satisfies the unitarity. 

\begin{table}
  \caption{The ground state properties of $^{16}$O:
    The bound single-particle levels (unit:MeV) and the r.m.s radius (unit:fm) for neutron and proton.}
  \label{table0}
  \begin{ruledtabular}
    \begin{tabular}{ccccc}
      & & Neutron & Proton &\\
      \colrule
      \multicolumn{5}{c}{hole states} \\
      & $s_{1/2}$ & -36.17 & -31.16 & \\
      & $p_{3/2}$ & -21.31 & -16.84 & \\
      & $p_{1/2}$ & -16.38 & -11.95 & \\
      \multicolumn{5}{c}{particle states} \\
      & $d_{5/2}$ &  -6.81 &  -2.95 & \\
      & $s_{1/2}$ &  -4.90 &  -1.43 & \\
      \\
      \multicolumn{5}{c}{r.m.s radius} \\
      & $\sqrt{\bra r^2\ket}$ & 2.41 & 2.46 & \\ 
    \end{tabular}
  \end{ruledtabular}
\end{table}

\begin{table}
  \caption{RPA configuration for the $J^{\pi}=2^+$ excitations of $^{16}$O}
  \label{table1}
  \begin{ruledtabular}
    \begin{tabular}{ccccccc}
      & $q=n$ or $p$ & $\alpha$ & $(lj)$ & $(lj)^{-1}$ & $\epsilon_\alpha$ & \\
      \colrule
       & &  1  &  $p_{3/2}$  &  $(p_{1/2})^{-1}$     & -16.38 & \\
       & &  2  &  $f_{5/2}$  &  $(p_{1/2})^{-1}$     & -16.38 & \\
       & &  3  &  $f_{5/2}$  &  $(p_{3/2})^{-1}$     & -21.31 & \\
       & $n$   &  4 &  $f_{7/2}$  &  $(p_{3/2})^{-1}$ & -21.31 & \\
       & &  5  &  $p_{1/2}$  &  $(p_{3/2})^{-1}$     & -21.31 & \\
       & &  6  &  $p_{3/2}$  &  $(p_{3/2})^{-1}$     & -21.31 & \\
       & &  7  &  $d_{3/2}$  &  $(s_{1/2})^{-1}$     & -36.17 & \\
       & &  8  &  $d_{5/2}$  &  $(s_{1/2})^{-1}$     & -36.17 & \\
      \\
      & &  9  &  $p_{3/2}$  &  $(p_{1/2})^{-1}$ & -11.95 & \\
      & & 10  &  $f_{5/2}$  &  $(p_{1/2})^{-1}$ & -11.95 & \\
      & & 11  &  $f_{5/2}$  &  $(p_{3/2})^{-1}$ & -16.84 & \\
  & $p$ & 12  &  $f_{7/2}$  &  $(p_{3/2})^{-1}$ & -16.84 & \\
      & & 13  &  $p_{1/2}$  &  $(p_{3/2})^{-1}$ & -16.84 & \\
      & & 14  &  $p_{3/2}$  &  $(p_{3/2})^{-1}$ & -16.84 & \\
      & & 15  &  $d_{3/2}$  &  $(s_{1/2})^{-1}$ & -31.16 & \\
      & & 16  &  $d_{5/2}$  &  $(s_{1/2})^{-1}$ & -31.16 & \\
    \end{tabular}
  \end{ruledtabular}
\end{table}

\subsection{Eigenphase shift decomposition of the RPA strength function and transition density}
\label{eigenphasedecom}
If we define the transition density $\vec{\bvec{\rho}}^{(\pm)\mathsf{T}}(r;E)$ as
an $\alpha_c$-dimensional vector as, 
\begin{eqnarray}
  \vec{\bvec{\rho}}^{(\pm)}(r;E)
  &\equiv&
  \sqrt{
    \frac{1}{\pi}
    \frac{2m}{\hbar^2}
  }
  \widehat{\bvec{\psi}}^{(\pm c)\mathsf{T}}(r;E)
  \vec{\bvec{\varphi}}(r)
  \label{trdenst},
\end{eqnarray}
it can be found that the strength function Eq.(\ref{RPAstr}) can be expressed as
\begin{eqnarray}
  &&
  S_F(E)
  \nonumber\\
  &&=
  \int dr
  f(r)
  \vec{\bvec{\rho}}^{(+)\dagger}(r;E^*)
  \int dr'
  f(r')  
  \vec{\bvec{\rho}}^{(+)}(r';E)
  \label{Sf-tdr}
  \\
  &&=
  \sum_{\alpha=1}^{\alpha_c}
  \int dr
  f(r)
  \rho^{(+\alpha)*}(r;E^*)
  \int dr'
  f(r')  
  \rho^{(+\alpha)}(r';E)
  \nonumber\\
  \label{Sf-tdr2}
\end{eqnarray}
by using the transition density, where $\rho^{(+\alpha)}$
is a component for $\alpha$ of the transition density vector $\vec{\bvec{\rho}}^{(+)}$. 
Therefore, the strength function on the real axis of the complex
excitation energy $E$ can be expressed as
\begin{eqnarray}
  S_F(E)
  &=&
  \sum_{\alpha=1}^{\alpha_c}
  \left|
  \int dr
  f(r)
  \rho^{(+\alpha)}(r;E)
  \right|^2
  \label{Sf-tdr3}
\end{eqnarray}
The relationship between $\vec{\bvec{\rho}}^{(+)}$
and $\vec{\bvec{\rho}}^{(-)}$ is given by
\begin{eqnarray}
  \vec{\bvec{\rho}}^{(+)}(r;E)
  &=&
  \bvec{\mathcal{S}}_{cc}(E)
  \vec{\bvec{\rho}}^{(-)}(r;E)
  \label{trdS-mat}
  \\
  &=&
  \bvec{\mathcal{S}}_{cc}(E)
  \vec{\bvec{\rho}}^{(+)*}(r;E^*)
  \label{trdS-mat2}
\end{eqnarray}
using the S-matrix.
\\
The S-matrix $\bvec{\mathcal{S}}_{cc}$, which is a unitary matrix, can be diagonalised using the unitary matrix defined
by $\bvec{\mathcal{U}}$ its eigenvectors as
\begin{eqnarray}
  \bvec{\mathcal{S}}_{cc}
  &=&
  \bvec{\mathcal{U}}
  \bvec{\mathcal{S}}_{\delta}
  \bvec{\mathcal{U}}^\dagger
  \label{diagSmat}
\end{eqnarray}
where $\bvec{\mathcal{S}}_{\delta}$ is a diagonal matrix with the complex eigenvalues of $\bvec{\mathcal{S}}_{cc}$ as matrix elements,
and since $|\det\bvec{\mathcal{S}}_{\delta}|=1$, $\bvec{\mathcal{S}}_{\delta}$ can be expressed as
\begin{eqnarray}
  \bvec{\mathcal{S}}_{\delta}
  =
  \begin{pmatrix}
    e^{2i\delta_1} &                  &        &                 &   \\
                & e^{2i\delta_2}      &        & \text{\huge{0}} &   \\
                &                  & \ddots &                 &   \\
                & \text{\huge{0}}  &        & \ddots          &            \\
                &                  &        &                 & e^{2i\delta_{\alpha_c}}
  \end{pmatrix}.
\end{eqnarray}
with $\delta_\alpha$ which is so-called as the eigenphase shift~\cite{}. 

If the ``eigenphase transition density''
$\vec{\bvec{\rho}}_{\delta}^{(\pm)}(r;E)$ is defined by the unitary transformation as
\begin{eqnarray}
  \vec{\bvec{\rho}}_{\delta}^{(\pm)}(r;E)
  \equiv
  \bvec{\mathcal{U}}^\dagger
  \vec{\bvec{\rho}}^{(\pm)}(r;E),
  \label{eigentrd}
\end{eqnarray}
Note that the vector component of $\vec{\bvec{\rho}}_{\delta}^{(\pm)}$, $\rho_{\delta}^{(+\alpha)}$,
is a linear combination of the vector component of $\vec{\bvec{\rho}}^{(\pm)}$, $\rho^{(\pm\alpha)}$,
and the coefficients are given by the matrix components of the unitary matrix $\bvec{\mathcal{U}}^\dagger$,
which diagonalises the S-matrix $\bvec{\mathcal{S}}_{cc}$.

The strength function with $\rho_{\delta}^{(+\alpha)}$ is just an expression
in which $\rho^{(+\alpha)}$ is replaced by $\rho_{\delta}^{(+\alpha)}$
in Eq.(\ref{Sf-tdr3}) as
\begin{eqnarray}
  S_F(E)
  &=&
  \sum_{\alpha=1}^{\alpha_c}
  \left|
  \int dr
  f(r)
  \rho_{\delta}^{(+\alpha)}(r;E)
  \right|^2
  \label{Sf-tdr4}
\end{eqnarray}
because the unitary transformation Eq.(\ref{eigentrd}) does not change the strength function. 

The T-matrix $\bvec{\mathcal{T}}_{cc}$ and K-matrix $\bvec{\mathcal{K}}_{cc}$
are defined by using the S-matrix $\bvec{\mathcal{S}}_{cc}$ as
\begin{eqnarray}
  \bvec{\mathcal{T}}_{cc}(E)
  &=&
  \frac{i}{2}
  \left(
  \bvec{\mathcal{S}}_{cc}(E)
  -\bvec{1}
  \right)
  \label{Tmat}
  \\
  \bvec{\mathcal{K}}_{cc}(E)
  &=&
  i
  \left(
  \bvec{\mathcal{S}}_{cc}(E)
  +\bvec{1}
  \right)^{-1}
  \left(
  \bvec{\mathcal{S}}_{cc}(E)
  -\bvec{1}
  \right)
  \label{Kmat},
\end{eqnarray}
the component for $\alpha$ of the diagonalized T-matrix $\bvec{\mathcal{T}}_{\delta}$ and K-matrix $\bvec{\mathcal{K}}_{\delta}$
are therefore expressed by using the eigenphase shift as
\begin{eqnarray}
  \mathcal{T}_{\delta_\alpha}(E)
  &=&
  -e^{i\delta_\alpha}\sin\delta_\alpha
  \label{Tmat-d}
  \\
  \mathcal{K}_{\delta_\alpha}(E)
  &=&
  -
  \tan\delta_\alpha.
  \label{Kmat-d}
\end{eqnarray}

\subsection{Eigenphase shift corresponding to RPA eigenstate}
\label{S1eigenphase}
In the RPA theory, collective excitation modes of nuclei, such as giant resonances, are thought
to be caused by the effects of residual interaction.
As given in Ref.\cite{JostRPA}, the Hamiltonian of the RPA equation as a simultaneous differential
equation in coordinate space representation is given by
\begin{eqnarray}
  \bvec{\mathcal{H}}
  &=&
  -\frac{\hbar^2}{2m}\del{^2}{r^2}
  \bvec{1}
  +
  \bvec{\mathcal{V}}_{MF}
  +
  \bvec{\mathcal{V}}_{res}
\end{eqnarray}
where the Hamiltonian is given by the $2N\times 2N$ matrix form, $\bvec{\mathcal{V}}_{MF}$ is the mean
field part which is given as the diagonal matrix, and $\bvec{\mathcal{V}}_{res}$ is the residual
interaction.
When the Hamiltonian is decomposed as
\begin{eqnarray}
  \bvec{\mathcal{H}}
  &=&
  \bvec{\mathcal{H}}_{MF}
  +
  \bvec{\mathcal{V}}_{res}
\end{eqnarray}
with
\begin{eqnarray}
  \bvec{\mathcal{H}}_{MF}
  &=&
  -\frac{\hbar^2}{2m}\del{^2}{r^2}
  \bvec{1}
  +
  \bvec{\mathcal{V}}_{MF}
\end{eqnarray}
and $\bvec{\mathcal{S}}_{MF}$ is defined as the S-matrix calculated using $\bvec{\mathcal{H}}_{MF}$.

$\bvec{\mathcal{S}}^{(1)}$ is defined by
\begin{eqnarray}
  \bvec{\mathcal{S}}^{(1)}  
  \equiv
  \bvec{\mathcal{S}}_{cc}
  \bvec{\mathcal{S}}_{MF}^\dagger
  \label{S1def}
\end{eqnarray}
The excited states of the system obtained by RPA theory are the eigenstates of
the Hamiltonian $\bvec{\mathcal{H}}$, which are created by the superposition of the basis defined by the
$\bvec{\mathcal{H}}_{MF}$ (particle-hole excited state configurations) with the effect of the residual interaction
$\bvec{\mathcal{V}}_{res}$. Based on this and the discussion in Ref. \cite{jost-class}, we can assume that the RPA eigenstates
are the $\bvec{\mathcal{S}}^{(1)}$ poles and that the strength function, by its definition, reflects the
$\bvec{\mathcal{S}}^{(1)}$ poles in its peak structure.
The poles of the $\bvec{\mathcal{K}}^{(1)}$ defined using $\bvec{\mathcal{S}}^{(1)}$ can be interpreted as
the RPA eigenvalues as a Sturm-Liouville problem and obtained on the real axis of energy as the energy when
the eigenphase shift $\delta^{(1)}$ becomes $\pi/2$, obtained by diagonalising $\bvec{\mathcal{S}}^{(1)}$. 
As stated in Ref.\cite{jost-class}, although the S- and K- matrices are related to each other, their poles are not
guaranteed to exist at the same time and to have a one-to-one correspondence.
However, the existence of poles of the K-matrix corresponding to the poles of the S-matrix is a condition
for the resonance property to be satisfied.

If $\vec{\bvec{\rho}}^{(+)}_{\delta^{(1)}}$ is defined as
\begin{eqnarray}
  \vec{\bvec{\rho}}^{(+)}_{\delta^{(1)}}
  \equiv
  \bvec{\mathcal{U}}^{(1)\dagger}
  \vec{\bvec{\rho}}^{(+)}
  \label{trd-S1}
\end{eqnarray}
using $\bvec{\mathcal{U}}^{(1)\dagger}$, which diagonalises $\bvec{\mathcal{S}}^{(1)}$ as,
\begin{eqnarray}
  \bvec{\mathcal{S}}^{(1)}
  &=&
  \bvec{\mathcal{U}}^{(1)}
  \bvec{\mathcal{S}}_{\delta}^{(1)}
  \bvec{\mathcal{U}}^{(1)\dagger}
  \label{diagS1}
\end{eqnarray}
since $\bvec{\mathcal{U}}^{(1)}$ is a unitary matrix,
the RPA strength function which is represented by Eq.(\ref{Sf-tdr})
can be expressed as
\begin{eqnarray}
  &&
  S_F(E)
  \nonumber\\
  &&=
  \int dr
  f(r)
  \vec{\bvec{\rho}}^{(+)\dagger}_{\delta^{(1)}}(r;E^*)
  \int dr'
  f(r')  
  \vec{\bvec{\rho}}^{(+)}_{\delta^{(1)}}(r';E)
  \label{Sf-tdr-S1}
\end{eqnarray}

\subsection{Isoscalar and Isovector strength}
\label{isivstr}
So far, for simplicity, the isospin dependence has not been specified, but the strength function,
which clearly shows the isospin dependence, is expressed as
\begin{eqnarray}
  &&
  S_{F,\tau\tau'}(E)
  \nonumber\\
  &&=
  \int dr
  f(r)
  \vec{\bvec{\rho}}^{(+)\dagger}_{\tau}(r;E^*)
  \int dr'
  f(r')  
  \vec{\bvec{\rho}}^{(+)}_{\tau'}(r';E)
  \label{Sf-tdrqq1}
\end{eqnarray}
with
\begin{eqnarray}
  \vec{\bvec{\rho}}^{(\pm)}_{\tau}(r;E)
  &\equiv&
  \sqrt{
    \frac{1}{\pi}
    \frac{2m}{\hbar^2}
  }
  \widehat{\bvec{\psi}}^{(\pm c)\mathsf{T}}(r;E)
  \vec{\bvec{\varphi}}_{\tau}(r).
  \label{trdenst-q}
\end{eqnarray}
where $\vec{\bvec{\varphi}}_{\tau}$ for $\tau=n,p$ are a $2N$-dimensional vector defined by
the hole-state wavefunction $\varphi_\alpha^{(q=n,p)}$ as,
\begin{eqnarray}
  \vec{\bvec{\varphi}}_{n}
  =
  \begin{pmatrix}
    \begin{pmatrix}
      \varphi_1^{(n)} \\
      \vdots\\
      \varphi_{N_n}^{(n)}
    \end{pmatrix}
    \\
    \begin{pmatrix}
      \varphi_1^{(n)} \\
      \vdots \\
      \varphi_{N_n}^{(n)}
    \end{pmatrix}
    \\
    \begin{pmatrix}
      0\\
      \vdots \\
      0
    \end{pmatrix}
    \\
    \begin{pmatrix}
      0\\
      \vdots \\
      0
    \end{pmatrix}
  \end{pmatrix},
  \hspace{10pt}
  \vec{\bvec{\varphi}}_{p}
  =
  \begin{pmatrix}
    \begin{pmatrix}
      0\\
      \vdots \\
      0
    \end{pmatrix}
    \\
    \begin{pmatrix}
      0\\
      \vdots \\
      0
    \end{pmatrix}
    \\
    \begin{pmatrix}
      \varphi_1^{(p)} \\
      \vdots\\
      \varphi_{N_n}^{(p)}
    \end{pmatrix}
    \\
    \begin{pmatrix}
      \varphi_1^{(p)} \\
      \vdots \\
      \varphi_{N_n}^{(p)}
    \end{pmatrix}
  \end{pmatrix},
\end{eqnarray}
and the ones for $\tau=0$(isoscalar) and $\tau=1$(isovector) are defined by
\begin{eqnarray}
  \vec{\bvec{\varphi}}_{0}
  &\equiv&
  \vec{\bvec{\varphi}}_{n}+\vec{\bvec{\varphi}}_{p}
  =
  \begin{pmatrix}
    \begin{pmatrix}
      \varphi_1^{(n)} \\
      \vdots\\
      \varphi_{N_n}^{(n)}
    \end{pmatrix}
    \\
    \begin{pmatrix}
      \varphi_1^{(n)} \\
      \vdots \\
      \varphi_{N_n}^{(n)}
    \end{pmatrix}
    \\
    \begin{pmatrix}
      \varphi_1^{(p)} \\
      \vdots\\
      \varphi_{N_p}^{(p)}
    \end{pmatrix}
    \\
    \begin{pmatrix}
      \varphi_1^{(p)} \\
      \vdots \\
      \varphi_{N_p}^{(p)}
    \end{pmatrix}
  \end{pmatrix}
  \label{holevec-is}
\end{eqnarray}
and
\begin{eqnarray}
  \vec{\bvec{\varphi}}_{1}
  &\equiv&
  \vec{\bvec{\varphi}}_{n}-\vec{\bvec{\varphi}}_{p}
  =
  \begin{pmatrix}
    \begin{pmatrix}
      \varphi_1^{(n)} \\
      \vdots\\
      \varphi_{N_n}^{(n)}
    \end{pmatrix}
    \\
    \begin{pmatrix}
      \varphi_1^{(n)} \\
      \vdots \\
      \varphi_{N_n}^{(n)}
    \end{pmatrix}
    \\
    -
    \begin{pmatrix}
      \varphi_1^{(p)} \\
      \vdots\\
      \varphi_{N_p}^{(p)}
    \end{pmatrix}
    \\
    -
    \begin{pmatrix}
      \varphi_1^{(p)} \\
      \vdots \\
      \varphi_{N_p}^{(p)}
    \end{pmatrix}
  \end{pmatrix}
  \label{holevec-iv}
\end{eqnarray}

\subsection{Decomposition of Isoscalar (IS) and Isovector (IV) modes in the absence of Coulomb interaction}
\label{noclmstr}
Decompose $\vec{\bvec{\rho}}^{(\pm)}_{\tau}(r;E)$,
given as an $\alpha_c$-dimensional vector in Eq.(\ref{trdenst-q}), as
\begin{eqnarray}
  \vec{\bvec{\rho}}^{(\pm)}_{\tau}
  &=&
  \begin{pmatrix}
    \vec{\bvec{\rho}}^{(\pm 1)}_{\tau} \\
    \vec{\bvec{\rho}}^{(\pm 2)}_{\tau} \\
    \vdots\\
    \vec{\bvec{\rho}}^{(\pm \alpha_c)}_{\tau}
  \end{pmatrix}
  =
  \begin{pmatrix}
    \vec{\bvec{\rho}}^{(\pm n)}_{\tau} \\
    \vec{\bvec{\rho}}^{(\pm p)}_{\tau}
  \end{pmatrix}
\end{eqnarray}
where $\vec{\bvec{\rho}}^{(\pm q)}_{\tau}$ (for $q=n$ or $p$) is the
$\alpha_c^{(q)}$-dimensional vector which is defined by
\begin{eqnarray}
  \vec{\bvec{\rho}}^{(\pm q)}_{\tau}
  &=&
  \begin{pmatrix}
    \vec{\bvec{\rho}}^{(\pm 1)}_{\tau} \\
    \vec{\bvec{\rho}}^{(\pm 2)}_{\tau} \\
    \vdots\\
    \vec{\bvec{\rho}}^{(\pm \alpha_c^{(q)})}_{\tau}
  \end{pmatrix}
\end{eqnarray}
with $\alpha_c=\alpha_c^{(n)}+\alpha_c^{(p)}$.
$\bvec{\mathcal{S}}_{cc}$, given as an $\alpha_c\times\alpha_c$ matrix,
is similarly decomposed as
\begin{eqnarray}
  \bvec{\mathcal{S}}_{cc}
  =
  \begin{pmatrix}
    \bvec{\mathcal{S}}_{cc}^{nn} & \bvec{\mathcal{S}}_{cc}^{np} \\
    \bvec{\mathcal{S}}_{cc}^{pn} & \bvec{\mathcal{S}}_{cc}^{pp} 
  \end{pmatrix}
\end{eqnarray}
where $\bvec{\mathcal{S}}_{cc}^{nn}$, $\bvec{\mathcal{S}}_{cc}^{np}$,
$\bvec{\mathcal{S}}_{cc}^{pn}$ and $\bvec{\mathcal{S}}_{cc}^{pp}$ are
$\alpha_c^{(n)}\times\alpha_c^{(n)}$, $\alpha_c^{(n)}\times\alpha_c^{(p)}$,
$\alpha_c^{(p)}\times\alpha_c^{(n)}$ and $\alpha_c^{(p)}\times\alpha_c^{(p)}$
matrices, respectively. 

Since $\bvec{\mathcal{S}}_{cc}$ is defined as in Eq.(\ref{Sccdef})
using $\bvec{\mathcal{S}}_{11}$ and $\bvec{\mathcal{S}}_{11}$ is expressed
as in Eq.(\ref{S11def}), we obtain
\begin{eqnarray}
  \bvec{\mathcal{S}}_{cc}^{nn}
  &=&
  \left[
    \left(\bvec{\mathcal{M}}^{(+)}_{nn}
    -
    \bvec{\mathcal{M}}^{(+)}_{np}
    \bvec{\mathcal{M}}^{(+)-1}_{pp}
    \bvec{\mathcal{M}}^{(+)}_{pn}\right)^{-1}
    \right.
    \nonumber\\
    &&
    \left.
    \left(\bvec{\mathcal{M}}^{(-)}_{nn}
    -
    \bvec{\mathcal{M}}^{(+)}_{np}
    \bvec{\mathcal{M}}^{(+)-1}_{pp}
    \bvec{\mathcal{M}}^{(-)}_{pn}\right)
    \right]_{cc}
  \\
  \bvec{\mathcal{S}}_{cc}^{np}
  &=&
  \left[
    \left(\bvec{\mathcal{M}}^{(+)}_{nn}
    -
    \bvec{\mathcal{M}}^{(+)}_{np}
    \bvec{\mathcal{M}}^{(+)-1}_{pp}
    \bvec{\mathcal{M}}^{(+)}_{pn}\right)^{-1}
    \right.
    \nonumber\\
    &&
    \left.
    \left(\bvec{\mathcal{M}}^{(-)}_{np}
    -
    \bvec{\mathcal{M}}^{(+)}_{np}
    \bvec{\mathcal{M}}^{(+)-1}_{pp}
    \bvec{\mathcal{M}}^{(-)}_{pp}\right)
    \right]_{cc}
  \\
  \bvec{\mathcal{S}}_{cc}^{pn}
  &=&
  \left[
    \left(\bvec{\mathcal{M}}^{(+)}_{pp}
    -
    \bvec{\mathcal{M}}^{(+)}_{pn}
    \bvec{\mathcal{M}}^{(+)-1}_{nn}
    \bvec{\mathcal{M}}^{(+)}_{np}\right)^{-1}
    \right.
    \nonumber\\
    &&
    \left.
    \left(\bvec{\mathcal{M}}^{(-)}_{pn}
    -
    \bvec{\mathcal{M}}^{(+)}_{pn}
    \bvec{\mathcal{M}}^{(+)-1}_{nn}
    \bvec{\mathcal{M}}^{(-)}_{nn}\right)
    \right]_{cc}
  \\
  \bvec{\mathcal{S}}_{cc}^{pp}
  &=&
  \left[
    \left(\bvec{\mathcal{M}}^{(+)}_{pp}
    -
    \bvec{\mathcal{M}}^{(+)}_{pn}
    \bvec{\mathcal{M}}^{(+)-1}_{nn}
    \bvec{\mathcal{M}}^{(+)}_{np}\right)^{-1}
    \right.
    \nonumber\\
    &&
    \left.
    \left(\bvec{\mathcal{M}}^{(-)}_{pp}
    -
    \bvec{\mathcal{M}}^{(+)}_{pn}
    \bvec{\mathcal{M}}^{(+)-1}_{nn}
    \bvec{\mathcal{M}}^{(-)}_{np}\right)
    \right]_{cc}
\end{eqnarray}
where $\bvec{\mathcal{M}}^{(\pm)}_{qq'}$ is the block matrices of
$\bvec{\mathcal{M}}^{(\pm)}$ which is defined by
\begin{eqnarray}
  \bvec{\mathcal{M}}^{(\pm)}
  \equiv
  \bvec{\mathcal{K}}_1^{\frac{1}{2}}
  \left(\bvec{\mathcal{J}}^{(\pm)}_{11}
  -
  \bvec{\mathcal{J}}^{(+)}_{12}
  \bvec{\mathcal{J}}^{(+)-1}_{22}
  \bvec{\mathcal{J}}^{(\pm)}_{21}
  \right)
  \bvec{\mathcal{K}}_1^{-\frac{1}{2}}.
\end{eqnarray}
The complex energy satisfying $\det\bvec{\mathcal{M}}^{(+)}=0$ gives
the poles of $\bvec{\mathcal{S}}_{cc}$, and $\det\bvec{\mathcal{M}}^{(+)}$
can also be expressed as
\begin{eqnarray}
  &&
  \det\bvec{\mathcal{M}}^{(+)}
  \nonumber\\
  &&=
  \det
  \bvec{\mathcal{M}}^{(+)}_{pp}
  \det
  \left(\bvec{\mathcal{M}}^{(+)}_{nn}
  -
  \bvec{\mathcal{M}}^{(+)}_{np}
  \bvec{\mathcal{M}}^{(+)-1}_{pp}
  \bvec{\mathcal{M}}^{(+)}_{pn}
  \right)
  \label{detM1}
  \\
  &&=
  \det
  \bvec{\mathcal{M}}^{(+)}_{nn}
  \det
  \left(\bvec{\mathcal{M}}^{(+)}_{pp}
  -
  \bvec{\mathcal{M}}^{(+)}_{pn}
  \bvec{\mathcal{M}}^{(+)-1}_{nn}
  \bvec{\mathcal{M}}^{(+)}_{np}
  \right)
  \label{detM2}
\end{eqnarray}

When there is no Coulomb interaction, the isospin symmetry leads
\begin{eqnarray}
  \vec{\bvec{\rho}}^{(\pm n)}_{n}
  &=&
  \vec{\bvec{\rho}}^{(\pm p)}_{p}
  \\
  \vec{\bvec{\rho}}^{(\pm n)}_{p}
  &=&
  \vec{\bvec{\rho}}^{(\pm p)}_{n}
  \\
  \bvec{\mathcal{M}}^{(+)}_{nn}
  &=&
  \bvec{\mathcal{M}}^{(+)}_{pp}
  \\
  \bvec{\mathcal{M}}^{(+)}_{np}
  &=&
  \bvec{\mathcal{M}}^{(+)}_{pn}
\end{eqnarray}
therefore, we have the relation for the isoscalar and isovector
transition density as
\begin{eqnarray}
  \vec{\bvec{\rho}}^{(\pm n)}_{0}
  &=&
  \vec{\bvec{\rho}}^{(\pm p)}_{0}
  \\
  \vec{\bvec{\rho}}^{(\pm n)}_{1}
  &=&
  -
  \vec{\bvec{\rho}}^{(\pm p)}_{1}.
\end{eqnarray}
Using these relations, we can obtain
\begin{eqnarray}
  \vec{\bvec{\rho}}^{(+q)}_0(r;E)
  &=&
  \left(
  \bvec{\mathcal{S}}_{cc}^{nn}(E)
  +
  \bvec{\mathcal{S}}_{cc}^{np}(E)
  \right)
  \vec{\bvec{\rho}}^{(-q)}_0(r;E)
  \label{trdS-mat_noclm1}
  \\
  \vec{\bvec{\rho}}^{(+q)}_1(r;E)
  &=&
  \left(
  \bvec{\mathcal{S}}_{cc}^{nn}(E)
  -
  \bvec{\mathcal{S}}_{cc}^{np}(E)
  \right)
  \vec{\bvec{\rho}}^{(-q)}_1(r;E)
  \label{trdS-mat_noclm2}
\end{eqnarray}
and also
\begin{eqnarray}
  &&
  \bvec{\mathcal{S}}_{cc}^{nn}\pm\bvec{\mathcal{S}}_{cc}^{np}
  \nonumber\\
  &&=
  \left[
    \left(
    \bvec{\mathcal{M}}^{(+)}_{nn}
    \pm
    \bvec{\mathcal{M}}^{(+)}_{np}
    \right)^{-1}
    \left(
    \bvec{\mathcal{M}}^{(-)}_{nn}
    \pm
    \bvec{\mathcal{M}}^{(-)}_{np}
    \right)
  \right]_{cc}.
\end{eqnarray}
The poles of $\bvec{\mathcal{S}}_{cc}^{nn}\pm\bvec{\mathcal{S}}_{cc}^{np}$
can be represented by the zeros of 
\begin{eqnarray}
  \det
    \left(
    \bvec{\mathcal{M}}^{(+)}_{nn}
    \pm
    \bvec{\mathcal{M}}^{(+)}_{np}
    \right)
    =0
    \label{Spole-ISIV}
\end{eqnarray}
on the complex energy plane.

Expressing the isospin dependence explicitly, 
Eqs.(\ref{trd-S1}) and (\ref{diagS1}) are expressed in block matrix form as
\begin{eqnarray}
  \begin{pmatrix}
    \vec{\bvec{\rho}}^{(+n)}_{\delta^{(1\alpha)},\tau} \\
    \vec{\bvec{\rho}}^{(+p)}_{\delta^{(1\beta)},\tau}
  \end{pmatrix}
  =
  \begin{pmatrix}
    \bvec{\mathcal{U}}^{(1)\dagger}_{n\alpha} & \bvec{\mathcal{U}}^{(1)\dagger}_{p\alpha} \\
    \bvec{\mathcal{U}}^{(1)\dagger}_{n\beta} & \bvec{\mathcal{U}}^{(1)\dagger}_{p\beta}
  \end{pmatrix}
  \begin{pmatrix}
    \vec{\bvec{\rho}}^{(+n)}_{\tau} \\
    \vec{\bvec{\rho}}^{(+p)}_{\tau}
  \end{pmatrix}
  \label{trd-S1-2d}
\end{eqnarray}
and
\begin{eqnarray}
  &&
  \begin{pmatrix}
    \bvec{\mathcal{S}}^{(1)}_{nn} & \bvec{\mathcal{S}}^{(1)}_{np} \\
    \bvec{\mathcal{S}}^{(1)}_{pn} & \bvec{\mathcal{S}}^{(1)}_{pp}
  \end{pmatrix}
  \nonumber\\
  &&=
  \begin{pmatrix}
    \bvec{\mathcal{U}}^{(1)}_{n\alpha} & \bvec{\mathcal{U}}^{(1)}_{n\beta} \\
    \bvec{\mathcal{U}}^{(1)}_{p\alpha} & \bvec{\mathcal{U}}^{(1)}_{p\beta}
  \end{pmatrix}
  \begin{pmatrix}
    \bvec{\mathcal{S}}_{\delta}^{(1\alpha)} & 0 \\
    0 & \bvec{\mathcal{S}}_{\delta}^{(1\beta)}
  \end{pmatrix}
  \begin{pmatrix}
    \bvec{\mathcal{U}}^{(1)\dagger}_{n\alpha} & \bvec{\mathcal{U}}^{(1)\dagger}_{p\alpha} \\
    \bvec{\mathcal{U}}^{(1)\dagger}_{n\beta} & \bvec{\mathcal{U}}^{(1)\dagger}_{p\beta}
  \end{pmatrix}
  \label{diagS1-2by2}.
\end{eqnarray}
When there is no Coulomb interaction, we can obtain the isoscalar and isovector
eigenphase transition density for $\bvec{\mathcal{S}}^{(1)}$ as
\begin{eqnarray}
  \vec{\bvec{\rho}}^{(+n)}_{\delta^{(1\alpha)},0}
  &=&
  \left(
  \bvec{\mathcal{U}}^{(1)\dagger}_{n\alpha}+\bvec{\mathcal{U}}^{(1)\dagger}_{p\alpha}
  \right)
  \vec{\bvec{\rho}}^{(+n)}_{0}
  \\
  \vec{\bvec{\rho}}^{(+n)}_{\delta^{(1\alpha)},1}
  &=&
  \left(
  \bvec{\mathcal{U}}^{(1)\dagger}_{n\alpha}-\bvec{\mathcal{U}}^{(1)\dagger}_{p\alpha}
  \right)
  \vec{\bvec{\rho}}^{(+n)}_{1}
\end{eqnarray}
from Eq.(\ref{trd-S1-2d}), and also we can obtain
\begin{eqnarray}
  \bvec{\mathcal{S}}^{(1)}_{nn}\pm\bvec{\mathcal{S}}^{(1)}_{np}
  =
  \bvec{\mathcal{U}}^{(1)}_{n\alpha}
  \bvec{\mathcal{S}}_{\delta}^{(1\alpha)}
  \left(
  \bvec{\mathcal{U}}^{(1)\dagger}_{n\alpha}\pm\bvec{\mathcal{U}}^{(1)\dagger}_{p\alpha}
  \right)
  \label{S1ISIV}
\end{eqnarray}
from Eq.(\ref{diagS1-2by2}).
These results show that when diagonalising $\bvec{\mathcal{S}}^{(1)}$,
the eigenphase shift $\delta^{(1\alpha)}$ with $\bvec{\mathcal{U}}^{(1)\dagger}_{n\alpha}=\bvec{\mathcal{U}}^{(1)\dagger}_{p\alpha}$
gives the Isoscalar mode and the eigenphase shift $\delta^{(1\alpha)}$ with
$\bvec{\mathcal{U}}^{(1)\dagger}_{n\alpha}=\bvec{\mathcal{U}}^{(1)\dagger}_{p\alpha}$ gives the Isovector mode.
Each mode is obtained independently when there is no Coulomb interaction.
When a Coulomb interaction is taken into account, the isospin symmetry is broken, so the isoscalar and isovector modes are not independent
solutions to each other. However, from the discussion in the absence of the Coulomb interaction, the eigenphase shift obtained by
diagonalising $\bvec{\mathcal{S}}^{(1)}$ can be interpreted as corresponding to the poles of $\bvec{\mathcal{S}}^{(1)}$ which
are given by zeros of the Jost function on the complex energy plane. 

\begin{figure}[htbp]
\includegraphics[width=\linewidth]{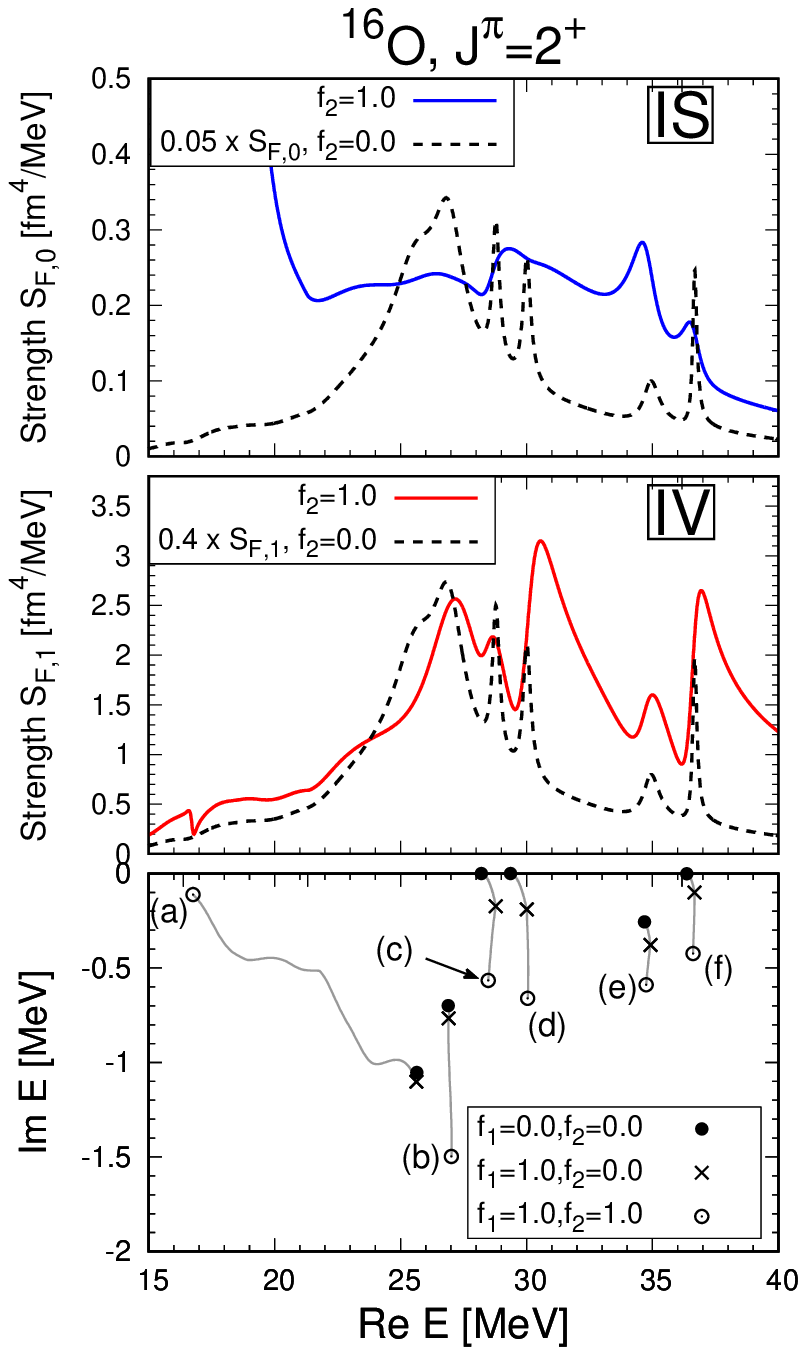}
\caption{(Color online) The isoscalar and isovector RPA strength functions for the quadrupole excitation of $^{16}$O
  were shown by the solid blue and red curves in the top and middle panels, respectively.
  The dashed line shows the RPA strength function calculated with $\kappa_{np}=0$,
  multiplied by factors $0.05$ and $0.4$ in the top and middle panels, respectively.
  The trajectories of poles (a)-(f) obtained by solving Eq.(\ref{detJ11}) varying the constant
  parameters $f_1$ and $f_2$ multiplied to the residual interaction are shown in the bottom
  panel. The open circle ($\circ$), cross ($\times$) and filled circle ($\bullet$) symbols show the position of poles which
  are calculated by the RPA (i.e., $f_1 = 1$, $f_2 = 1$), RPA with $\kappa_{np} = 0$ ($f_1 = 1$,
  $f_2 = 0$) and unperturbed($f_1 = 0$, $f_2 = 0$) solutions, respectively.(See Table.\ref{table2}.)
}
\label{srttrjO16L2}
\end{figure}

\section{Results and discussions}
\label{res-dis}

In the numerical calculations in this paper the same model and parameters as in
Ref.\cite{JostRPA} are used for the Woods-Saxon potential and the residual
interaction. 
The target nucleus was also chosen as a relatively light spherical nucleus,
$^{16}$O as before, in order to understand the details of the method and to
proceed with the analysis.
In this paper, the RPA quadrupole excitation is calculated by adopting $f(r)=r^2$
as the external field; the distinction between isoscalar(IS) and isovector(IV)
modes is given by the sign of the hole state vector, as given in Eqs. (\ref{holevec-is})
and (\ref{holevec-iv}).
The ground state properties, the single particle levels and r.m.s radius for neutron
and proton of $^{16}$O are shown in Table \ref{table0}. The configuration of the quadrupole excitations
of $^{16}$O and subscription $\alpha$ which is used to describe the matrix elements 
of the Jost function are shown in Table \ref{table1}.
A minor change regarding the numerical solution method from Ref.\cite{JostRPA},
but the Newton-Raphson method~\cite{Rakityansky,NRM} is adopted in this paper to solve
Eq.(\ref{detJ11}) and find the S-matrix poles on the complex energy plane.

\begin{table*}
  \caption{Values of the S-matrix poles shown in the bottom panel of Fig.\ref{srttrjO16L2}. ``Sheet'' is the Riemann sheet on which the poles are found.
    The energy regions given in parentheses are the branch-cut lines of analytical continuation on the real axis with the first Riemann sheet.
    The ``origin'' gives the configuration of the single-particle excitation of the poles, calculated with $f_1=f_2=0.0$.}
  \label{table2}
  \begin{ruledtabular}
    \begin{tabular}{cccccc}
      & & \multicolumn{3}{c}{S-matrix pole [MeV]} & \\
      \cline{3-5} 
      Sheet (branch-cut [MeV]) & No. & $f_1=f_2=1.0$ & $f_1=1.0, f_2=0.0$ & $f_1=f_2=0.0$ & origin \\
      && ($\circ$ in Fig.\ref{srttrjO16L2}) & ($\times$ in Fig.\ref{srttrjO16L2}) & ($\bullet$ in Fig.\ref{srttrjO16L2}) & \\
      \colrule
      Sheet 1 ($11.95 \leq E \leq 16.38$) & -- & -- & -- & -- & -- \\
      Sheet 2 ($16.38 \leq E \leq 16.84$)                  & (a) & $16.76-i 0.11$ & $25.62-i1.10$ & $25.65-i1.05$ & $\pi[f_{7/2}\otimes (p_{3/2})^{-1}]$ \\
      Sheet 3 ($16.84 \leq E \leq 21.31$) & -- & -- & -- & -- & -- \\
      \multirow{3}{*}{Sheet 4 ($21.31 \leq E \leq 31.16$)} & (b) & $27.02-i 1.50$ & $26.91-i0.77$ & $26.89-i0.70$ & $\nu[f_{7/2}\otimes (p_{3/2})^{-1}]$ \\
                                                           & (c) & $28.48-i 0.57$ & $28.78-i0.17$ & $28.21$ & $\pi[d_{5/2}\otimes (s_{1/2})^{-1}]$ \\
                                                           & (d) & $30.47-i 0.66$ & $30.01-i0.19$ & $29.36$ & $\nu[d_{5/2}\otimes (s_{1/2})^{-1}]$ \\
      Sheet 5 ($31.16 \leq E \leq 36.17$)                  & (e) & $34.75-i 0.59$ & $34.91-i0.37$ & $34.69 -i 0.26$ & $\pi[d_{3/2}\otimes (s_{1/2})^{-1}]$ \\
      Sheet 6 ($36.17 \leq E$)                             & (f) & $36.61-i 0.42$ & $36.67-i0.10$ & $36.36 -i 0.001$ & $\nu[d_{3/2}\otimes (s_{1/2})^{-1}]$ 
    \end{tabular}
  \end{ruledtabular}
\end{table*}

In the top and middle panels of Fig.\ref{srttrjO16L2}, the RPA isoscalar and
isovector strength functions are shown by the blue and red solid curves, respectively.
The dashed curve shows the RPA strength function calculated by ignoring the residual
interaction between neutrons and protons.
To compare the peak structures of the solid and dashed strength functions in the energy
region above 20 MeV, the dashed strength functions were multiplied by 0.05 (top panel)
and 0.4 (middle panel). 
The bottom panel of Fig.\ref{srttrjO16L2} shows the S-matrix poles and their trajectories
obtained by solving Eq.(\ref{detJ11}). The trajectories were obtained by varying the constants
multiplied by the residual interaction in the same way as done in Ref.\cite{JostRPA}.
Depending on the poles, the Riemann sheet to which the poles belong is different;
the values of the S-matrix poles and the Riemann sheets to which they belong are shown in
Table \ref{table2}. 

\begin{figure}[htbp]
\includegraphics[width=\linewidth]{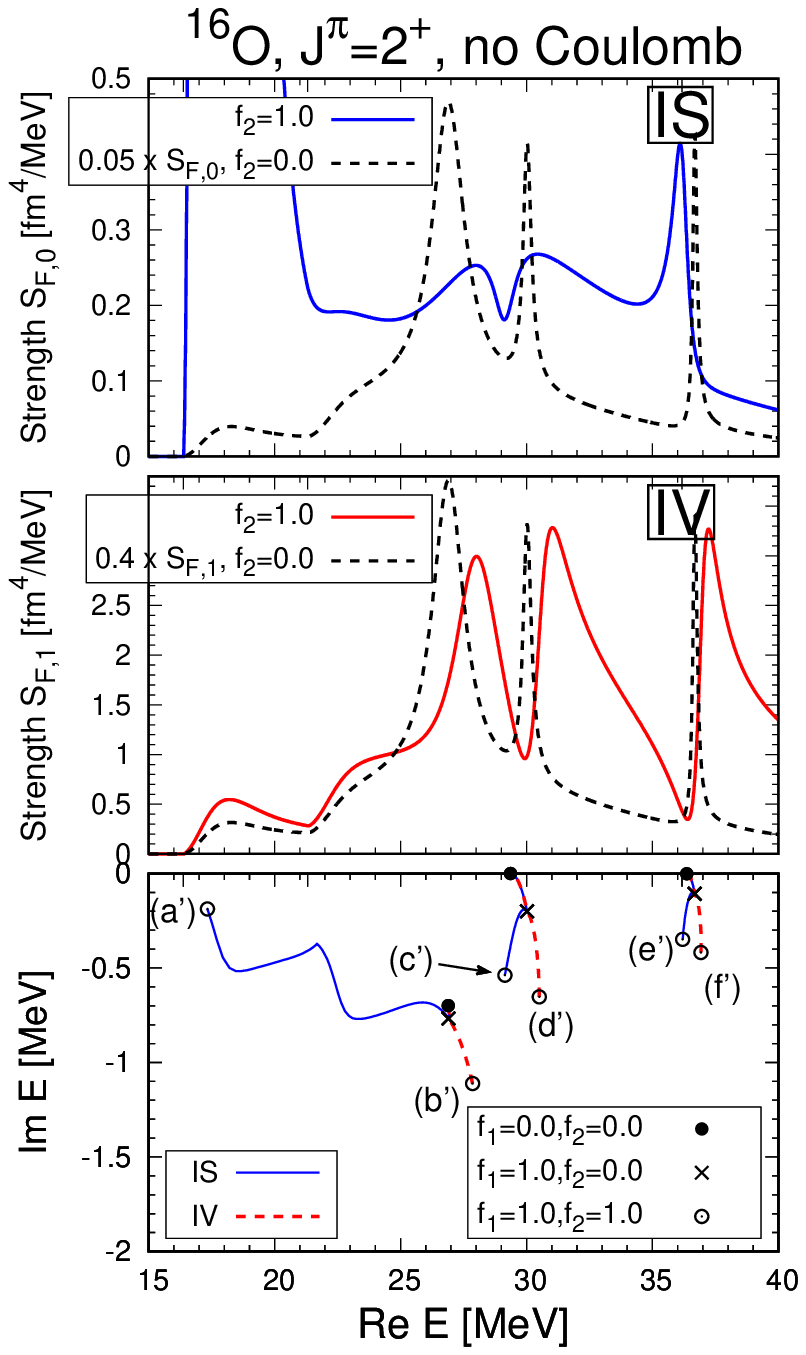}
\caption{(Color online) The same figure with Fig.\ref{srttrjO16L2}
  but calculated without Coulomb interaction. The trajectories of poles are
  calculated by solving Eq.(\ref{Spole-ISIV}), the positive and negative signs for
  the isoscalar and isovector modes, respectively. (See. Table.\ref{table3}.) 
}
\label{srttrjO16L2noCLM}
\end{figure}

Basically, isoscalar (IS) and isovector (IV) modes are excitation modes caused by neutron-proton
coupling. Therefore, there is no distinction between IS and IV modes in the strength function
indicated by the dashed curve in the upper and middle panels, and it can be easily confirmed
that the S-matrix poles marked with cross($\times$) symbols in the bottom panel correspond to
the peaks of the strength function indicated by the dashed curve.

However, in RPA calculations considering neutron-proton coupling, it is known that IS and IV modes
are mixed when the neutron and proton density distributions are different. (Mixing of IS and IV
modes occurs even in Z=N nuclei, such as $^{16}$O, due to isospin symmetry breaking by the presence
of the Coulomb interaction.)\cite{sagawa}.
It is therefore difficult to find a clear correspondence between the poles and the peaks of the RPA
strength function, except for a pole (a), which are clearly isolated from the other poles. 
\begin{table*}
  \caption{Almost the same table as Table \ref{table2} for Fig.\ref{srttrjO16L2noCLM}.
  The ``mode'' shown in the last column indicates the excitation mode (IS: Isoscalar or IV: Isovector) for each pole.}
  \label{table3}
  \begin{ruledtabular}
    \begin{tabular}{ccccccc}
      & & \multicolumn{3}{c}{S-matrix pole [MeV]} & &\\
      \cline{3-5} 
      Sheet (branch-cut [MeV]) & No. & $f_1=f_2=1.0$ & $f_1=1.0, f_2=0.0$ & $f_1=f_2=0.0$ & origin & mode\\
      && ($\circ$ in Fig.\ref{srttrjO16L2noCLM}) & ($\times$ in Fig.\ref{srttrjO16L2noCLM}) & ($\bullet$ in Fig.\ref{srttrjO16L2noCLM}) & &\\
      \colrule
      Sheet 1 ($16.38 \leq E \leq 21.31$)                  & (a') & $17.33-i 0.19$ & \multirow{2}{*}{$26.90-i0.77$} & \multirow{2}{*}{$26.89-i0.70$}    & \multirow{2}{*}{$[f_{7/2}\otimes (p_{3/2})^{-1}]$} & IS\\
      \multirow{3}{*}{Sheet 2 ($21.31 \leq E \leq 36.17$)} & (b') & $27.86-i 1.11$ & & &  & IV\\
                                                           & (c') & $29.14-i 0.54$ & \multirow{2}{*}{$30.02-i0.20$} & \multirow{2}{*}{$29.36$}          & \multirow{2}{*}{$[d_{5/2}\otimes (s_{1/2})^{-1}]$} & IS\\
                                                           & (d') & $30.50-i 0.65$ & & & & IV\\
      \multirow{2}{*}{Sheet 3 ($36.17 \leq E$)}            & (e') & $36.20-i 0.35$ & \multirow{2}{*}{$36.68-i0.11$} & \multirow{2}{*}{$36.36 -i 0.001$} & \multirow{2}{*}{$[d_{3/2}\otimes (s_{1/2})^{-1}]$} & IS\\
                                                           & (f') & $36.92-i 0.42$ & & & & IV
    \end{tabular}
  \end{ruledtabular}
\end{table*}
For $Z=N$ nuclei, the IS and IV modes can be obtained as independent solutions without mixing if the
Coulomb interaction is neglected. Fig.\ref{srttrjO16L2noCLM} is the same as Fig.\ref{srttrjO16L2} but
ignoring the Coulomb interaction. 
Using Eq.(\ref{Spole-ISIV}), the S-matrix poles can also be obtained independently for the IS and IV modes.
The results of Fig.\ref{srttrjO16L2noCLM} and Table \ref{table3} show that (a'), (c') and (e') are the
poles of the IS mode and (b'), (d') and (f') are the poles of the IV mode. (a') and (b'), (c') and (d'), (e')
and (f') are IS and IV modes, respectively, originating from the same state due to the effect of residual
interaction.
In other words, in the case of a pair of poles (a') of an IS mode and a pole (b') of an IV mode, for example,
the origin of both poles is a pole at $26.89-i 0.70$ MeV. Firstly, this becomes a pole existing at $26.90-i 0.77$
MeV due to the effect of residual interaction between homologous nucleons. This is then further split into an
independent IS-mode pole($17.33-i0.19$ MeV) and an IV-mode pole($27.86-i1.11$ MeV) by the effect of residual
interaction between neutrons and protons.

Comparing the positions of the S-matrix poles indicated by open circle symbols($\circ$) in
Fig.\ref{srttrjO16L2noCLM} and the peaks of the strength function represented by solid curves, it seems that
the real part of the S-matrix poles generally corresponds to the peaks of the strength function, except
the IS mode (c').
A comparison of the S-matrix poles in Figs.\ref{srttrjO16L2} and \ref{srttrjO16L2noCLM} shows that
there is a correspondence. 
In the presence of the Coulomb interaction, the origin pole separates into two poles due to isospin symmetry breaking,
as shown in Fig.\ref{srttrjO16L2} and Table \ref{table2}. In the example of the previous pair (a') and (b'),
the origin pole at $26.89-i 0.70$ MeV separates into $25.65-i1.05$ MeV and $26.89-i0.70$ MeV poles, which
become pole (a) and pole (b) respectively due to residual interaction. 
\begin{figure}[htbp]
\includegraphics[width=\linewidth]{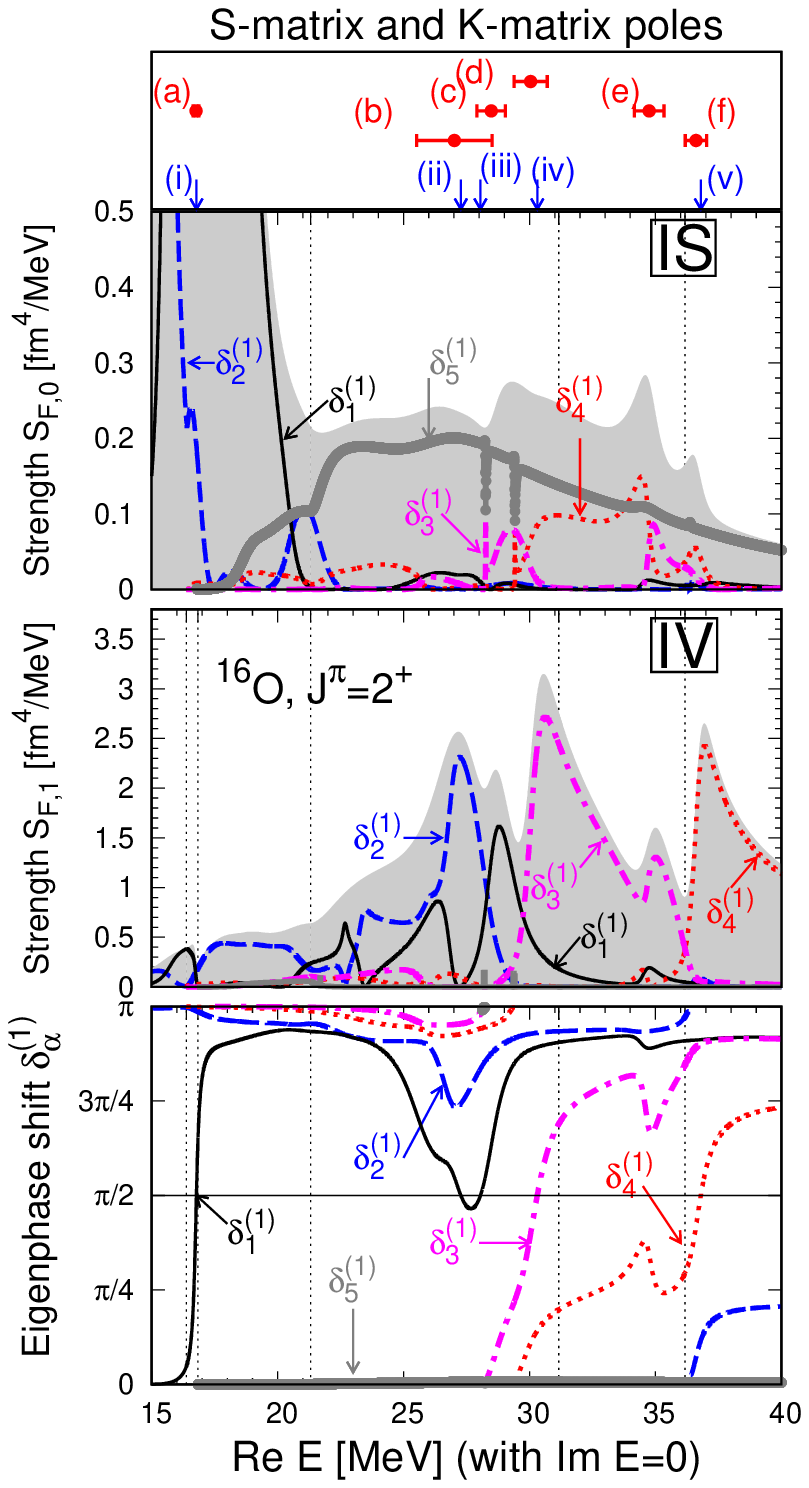}
\caption{(Color online)
  Eigenphase shift $\delta^{(1)}_\alpha$ obtained by diagonalising the
  $\bvec{\mathcal{S}}^{(1)}$ matrix
  using Eq.(\ref{diagS1}) (bottom panel) and decomposition of the RPA strength
  function by the eigenphase shift components (second and third panels).
  (a)-(f) in the upper figure are the S-matrix poles shown in Fig.\ref{srttrjO16L2}
  and Table \ref{table2}. The imaginary part of the poles is represented by error
  bars. 
  (i)-(v) show the positions of the K-matrix poles appearing on the real axis of energy.
  The energy of the K-matrix pole is equivalent to the energy when the eigenphase shift
  $\delta^{(1)}_\alpha$ crosses $\pi/2$. (See Table.\ref{table4}.)}
\label{srtphsO16L2}
\end{figure}
\begin{figure}[htbp]
\includegraphics[width=\linewidth]{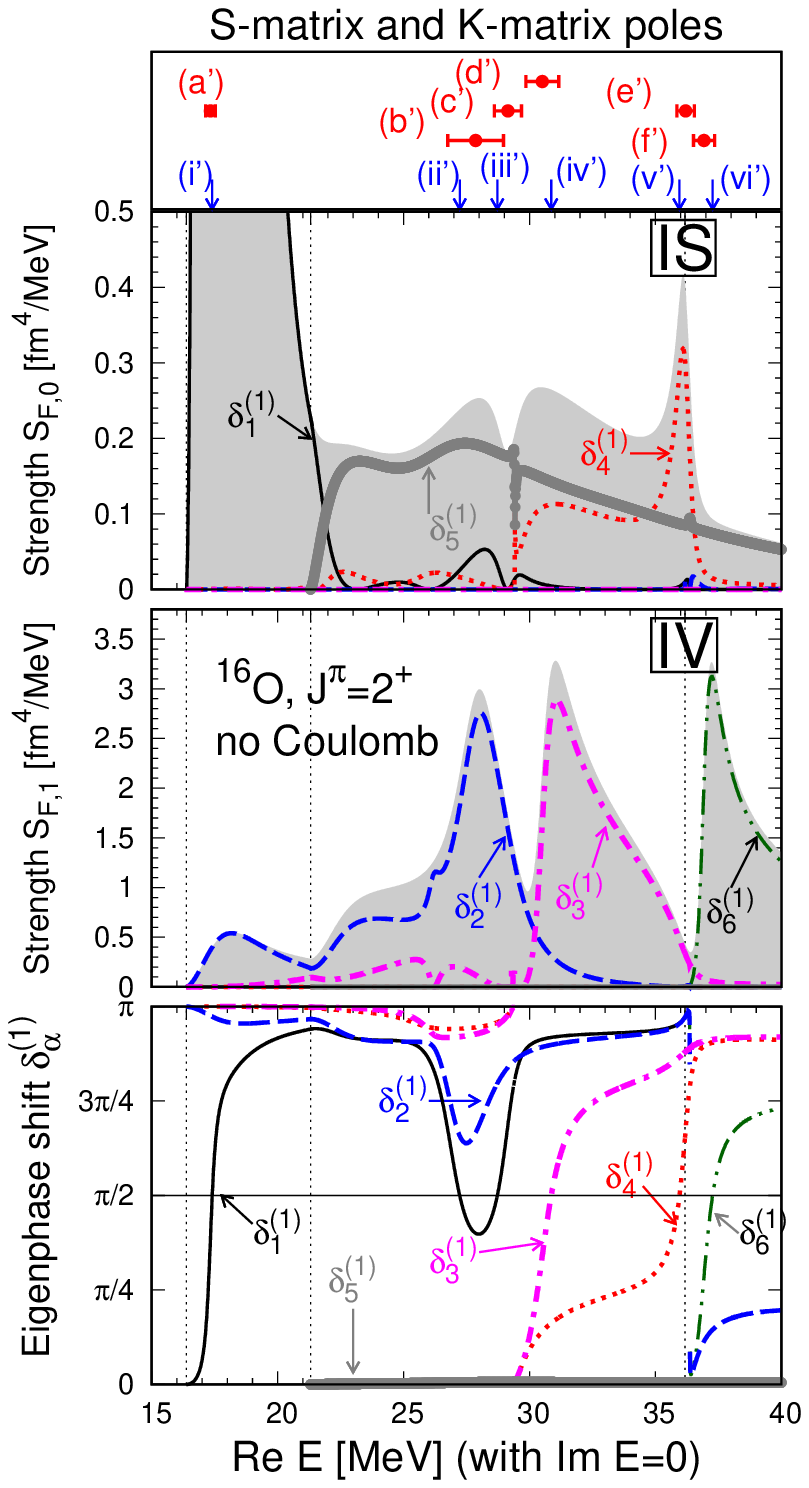}
\caption{(Color online)
  The same figure with Fig.\ref{srtphsO16L2} but calculated without Coulomb interaction.}
\label{srtphsO16L2noCLM}
\end{figure}

However, as discussed in Ref.\cite{jost-fano,jost-class}, it does not always guarantee that the position of
the real part of the S-matrix poles in general reflects or corresponds to the position of the peak of a physical
quantity such as the cross section or strength function.
This is because the Jost function is essentially a multivalued function of complex energy, so there can be energy shifts
caused by the nature of the multivalued function, and if several poles exist in close proximity to each other, there is a
superposition of the contributions of the individual poles. There are also background contributions arising from non-resonant
continuum state contributions, because interference effects, including Fano effects, also affect the peak structure.
Furthermore, there is a mixture of IS and IV modes, so it is not easy to derive conclusions about the correspondence between
the peak structure of the strength function and the S-matrix poles and their contributions.

The Gamow shell model\cite{gamow1,gamow2} and the complex scaling method\cite{CSM1,CSM2} exist as methods which can decompose the contribution
of the S-matrix poles in the strength function and cross section. Further extension of the method is underway to apply these
methods to our Jost function method. In this paper, we attempt the alternative way for the detailed analysis.

In the previous section of this paper, we derived the S-matrix which satisfies the unitarity associated with the description of
the RPA strength function within the framework of RPA theory using the Jost function.
The S-matrix which satisfies unitarity can be diagonalised using the unitary matrices as shown in Eq.(\ref{diagS1}),
which provides the eigenphase shifts.
Since the unitary transformation is an isometric transformation, the unitary matrix, which diagonalises the S-matrix, can
decompose the RPA intensity function into components for each eigenphase shift without changing its magnitude.
The poles of the K-matrix, defined using the S-matrix, appear on the real axis of energy and are equal to the energy at which
the eigenphase shift crosses $\pi/2$, referred to as the eigenvalue of the Sturm-Liouville problem.

\begin{figure}[htbp]
\includegraphics[width=\linewidth]{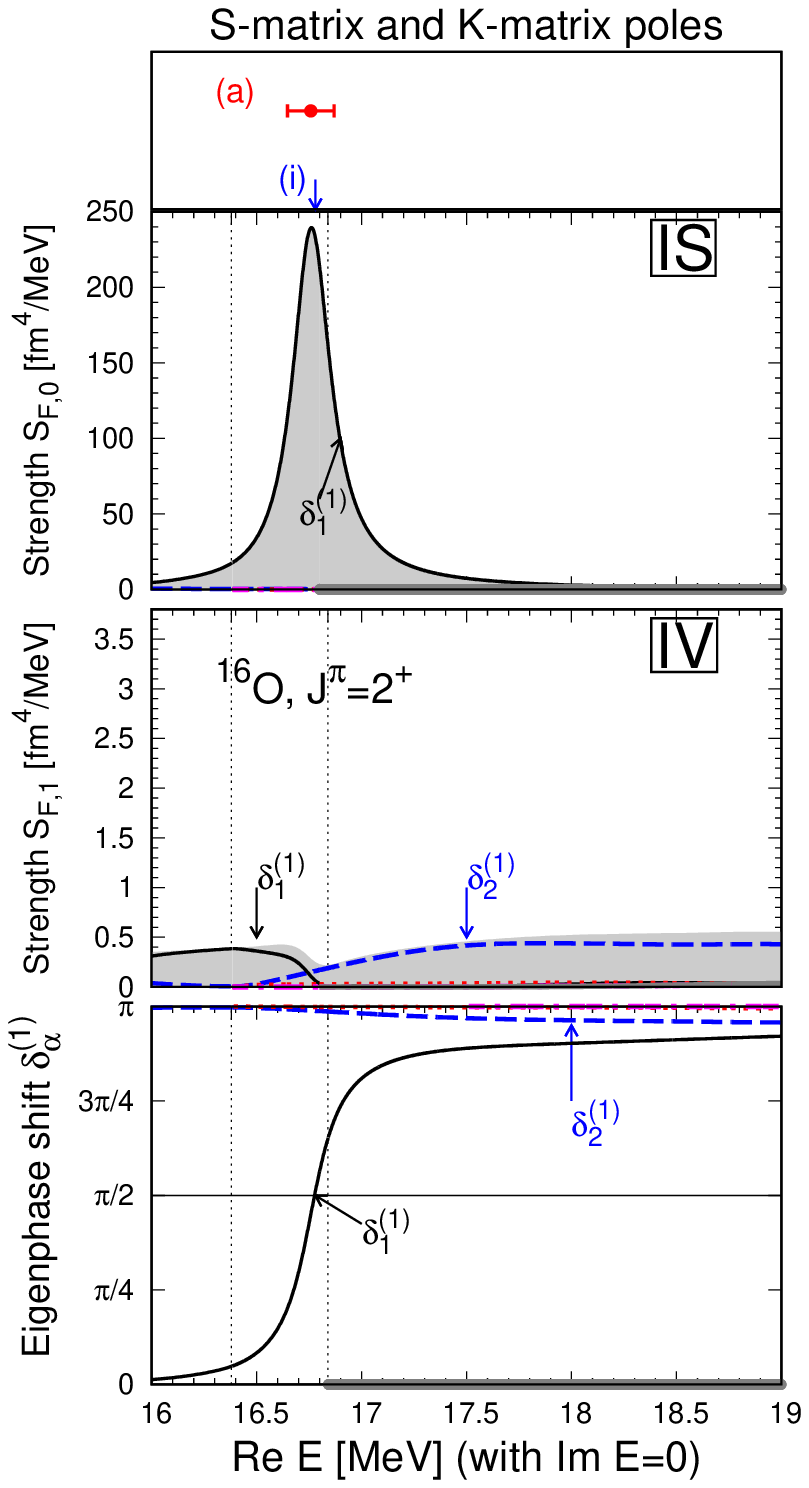}
\caption{(Color online)
  Magnified view of the energy region from $16$ MeV to $19$ MeV in Fig.\ref{srtphsO16L2}.}
\label{srtphsO16L2low}
\end{figure}
\begin{figure}[htbp]
\includegraphics[width=\linewidth]{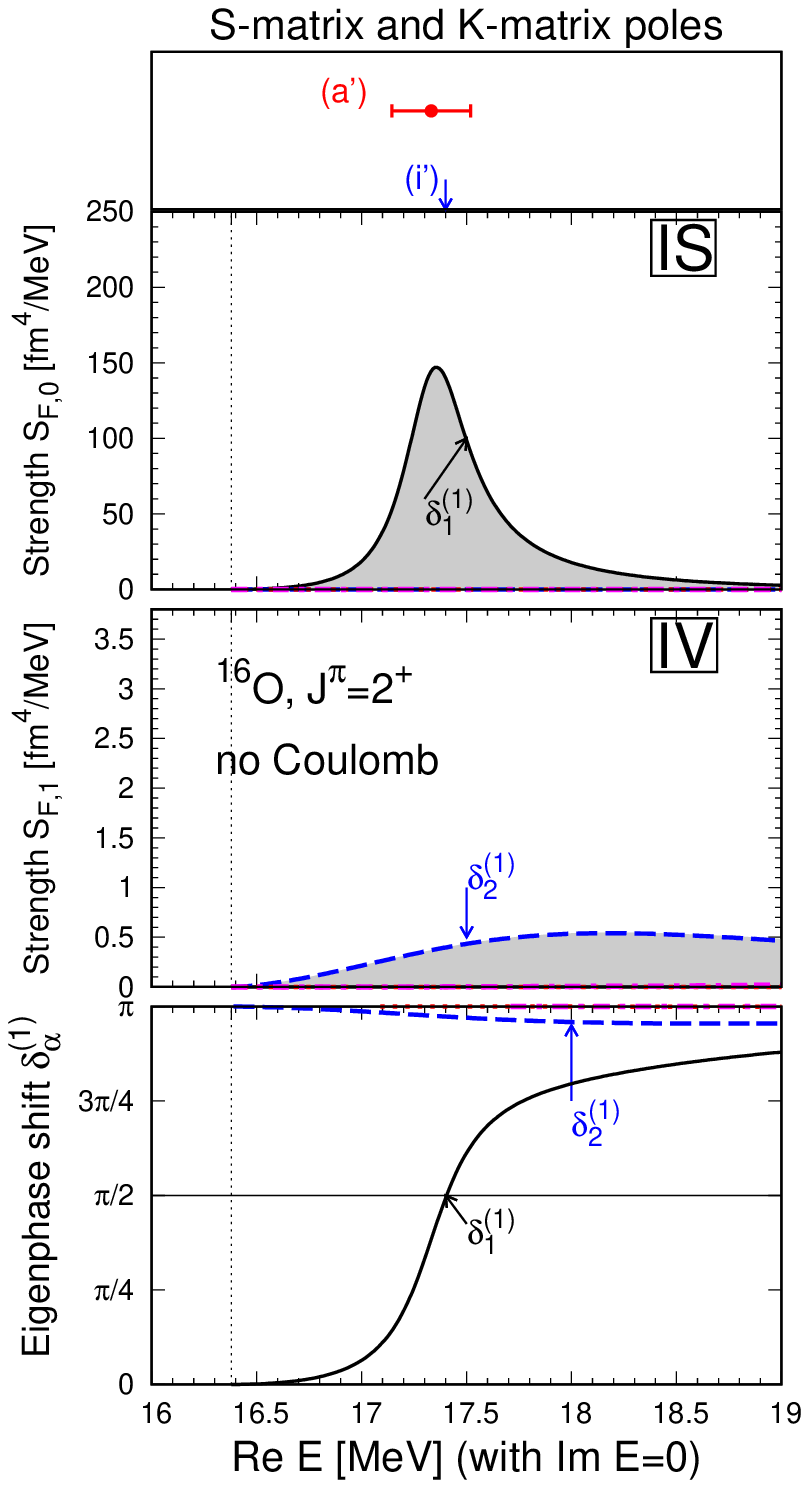}
\caption{(Color online)
  Magnified view of the energy region from $16$ MeV to $19$ MeV in Fig.\ref{srtphsO16L2noCLM}.}
\label{srtphsO16L2noCLMlow}
\end{figure}

Eigenphase shifts $\delta^{(1)}_\alpha$ obtained by diagonalising $\bvec{\mathcal{S}}^{(1)}$ are shown in the bottom panel of
Fig. \ref{srtphsO16L2}. The first and third rows of the figure show the strength function decomposed into the components giving
the eigenphase shifts using Eq.(\ref{Sf-tdr-S1}). In the top panel of Fig.\ref{srtphsO16L2}, the S-matrix poles (a)-(f)
are shown with their imaginary parts represented by error-bars, since it is known that the real part of the S-matrix pole is the
resonance energy and the imaginary part is the half-width if it is an isolated pole.
(i)-(v) show the location of the K-matrix poles on the real axis of energy.
As is clear from the derivation of the S-matrix in the previous section, the size of the S-matrix depends on the energy region
of the Riemann sheet shown in Table \ref{table2}.
For example, in the energy region of Sheet 2 ($16.38 \leq E \leq 16.84$ MeV),
the S-matrix is a $4\times 4$ matrix because $\alpha=1,2,9$ and $10$ are open in the
configuration channels given in Table \ref{table2}. In the case of Sheet 4
($21.31 \leq E \leq 31.16$ MeV), the S-matrix is a $4\times 4$ matrix because $\alpha=1-6,9-14$
channels are open.
Energy regions are indicated by dotted vertical lines in Fig. \ref{srtphsO16L2}. Despite the fact that the eigenphase shift
is obtained by diagonalising S-matrices of different sizes in each energy region, the eigenphase shift is obtained as a
function of energy, which is continuously connected at the boundaries of the energy region.
The components of the strength function per eigenphase shift are also obtained as a function of continuous energy at the
boundaries of the energy region, and the sum of all those components reproduces the total strength function (shown as grey-filled area).
Note that the ``total strength function'' is the same as the strength function which can be obtained by using the cRPA method~\cite{shlomo}. 
The same figure with Fig.\ref{srtphsO16L2} but calculated without Coulomb interaction is shown in Fig.\ref{srtphsO16L2noCLM}.
In order to show the first quadrupole exited state clearly, the magnified view of the energy region from $16$ MeV to $19$ MeV
of Figs.\ref{srtphsO16L2} and \ref{srtphsO16L2noCLM} are shown in Figs.\ref{srtphsO16L2low} and \ref{srtphsO16L2noCLMlow},
respectively. 

If the Coulomb interaction is neglected, the eigenphase shifts shown in Fig.\ref{srtphsO16L2noCLM} can be obtained
independently for the IS and IV modes using Eq.(\ref{S1ISIV}). As a result, it can be seen that $\delta^{(1)}_1$,
$\delta^{(1)}_4$ and $\delta^{(1)}_5$ are the eigenphase shifts of the IS mode, while $\delta^{(1)}_2$, $\delta^{(1)}_3$
and $\delta^{(1)}_6$ are those of the IV mode.
These results are also consistent with the results of the component decomposition of the strength function which
are shown above the eigenphase shifts.
The energy at which the eigenphase shift crosses $\pi/2$ is the K-matrix pole; it is known that if the S-matrix
pole is given by a single pole, the energy derivative $\frac{d\delta^{(1)}_\alpha}{dE}$ of the phase shift corresponds
to the imaginary part of the pole (resonance width) at the energy at which the phase shift crosses $\pi/2$.
It is therefore reasonable to consider (ii') shown in Fig.\ref{srtphsO16L2noCLM} as an independent K-matrix pole
which is not associated with the S-matrix pole.
For the K-matrix poles, except for (ii'), the correspondence with the S-matrix poles can be found from the
difference in energy positions and modes. The results of the correspondence are summarized in the lower half
of Table \ref{table4}.

\begin{table}
  \caption{Values of the K-matrix poles and corresponding eigenphase shifts shown in Figs. \ref{srtphsO16L2},
    \ref{srtphsO16L2low}, \ref{srtphsO16L2noCLM} and \ref{srtphsO16L2noCLMlow}.}
  \label{table4}
  \begin{ruledtabular}
    \begin{tabular}{ccccccc}
      & No.   & Eigenphase & K-mat. pole & Corr. & mode &\\
      &    &  & [MeV] & S-mat.  &&\\
      \colrule
      \multicolumn{5}{c}{with Coulomb} \\
      & (i)   & $\delta_1^{(1)}$ & $16.78$ & (a) & IS & \\
      & (ii)  & $\delta_1^{(1)}$ & $27.27$ & -- & ind. &\\
      & (iii) & $\delta_1^{(1)}$ & $28.05$ & (c) &&\\
      & (iv)  & $\delta_3^{(1)}$ & $30.31$ & (d) &&\\
      & (v)   & $\delta_4^{(1)}$ & $36.80$ & (e),(f) &&\\
      \multicolumn{5}{c}{without Coulomb} \\
      & (i')   & $\delta_1^{(1)}$ & $17.40$ & (a') & IS &\\
      & (ii')  & $\delta_1^{(1)}$ & $27.23$ & -- & ind. &\\
      & (iii') & $\delta_1^{(1)}$ & $28.72$ & (c') & IS &\\
      & (iv')  & $\delta_3^{(1)}$ & $30.86$ & (d') & IV &\\
      & (v')   & $\delta_4^{(1)}$ & $35.95$ & (e') & IS &\\
      & (vi')  & $\delta_6^{(1)}$ & $37.27$ & (f') & IV &\\
    \end{tabular}
  \end{ruledtabular}
\end{table}

As seen in Fig. \ref{srtphsO16L2noCLMlow}, the eigenphase shift $\delta^{(1)}_1$ gives a very large strength function
of the IS first quadrupole excited state around $E = 17.40$ MeV of the K-matrix pole (i'); the value of the imaginary part
of the S-matrix pole corresponding to $\delta^{(1)}_1$ (given in Table \ref{table3} shows that the width of this peak
is $0.38$ MeV ($= 0.19\times 2$).
In contrast, despite the presence of the corresponding S-matrix pole (c') for the K-matrix pole (iii'),
$\delta^{(1)}_5$ makes only a minor contribution to the strength function of the IS mode around the energy of (iii') $E=28.72$ MeV.
Near this energy, the main contribution is given by the eigenphase shift $\delta^{(1)}_5$ (non-resonant continuum state),
which has no K-matrix pole and is close to zero.
However, the basic shape of the strength function near the K-matrix pole (iii') is given by the strength function component
contributed by $\delta^{(1)}_1$. The shape of this strength function $\delta^{(1)}_1$ component is asymmetric, which means that
quantum interference effects such as the Fano effect may exist.
The Fano effect (or Fano resonance) is known as a special quantum interference effect that is universal to quantum
many-body systems and is known in atomic physics as the effect that causes autoionisation phenomena~\cite{fano,Chu,moiseyev,Desrier}.
As far as the RPA strength function is concerned, its contribution is very small and negligible compared to the first excited
state, but as the K-matrix pole (iii') is a state with a corresponding S-matrix pole (c'), it is expected to give a clear
contribution to quantities related to the scattering cross section such as the imaginary part of the T-matrix. However,
as our numerical calculations and analysis in this paper focus on the RPA strength function, we can only point out the
possible existence of the Fano effect.

The $\delta^{(1)}_2$ component in Fig.\ref{srtphsO16L2noCLM} is considered to correspond to the S-matrix pole (b')
because it gives the IV mode contribution from the strength function, but there is no corresponding K-matrix pole.
Therefore, the S-matrix pole (d') is considered to be an independent S-matrix pole which forms the first peak of
the IV mode around $E=27$ MeV, but without a corresponding K-matrix pole.

The eigenphase shifts $\delta^{(1)}_3$ and $\delta^{(1)}_6$ give the K-matrix poles (iv') and (vi') with corresponding
IV-mode S-matrix poles (d') and(f') respectively. These are then the dominant components of the peaks of the IV strength
function near the respective K-matrix poles.

The K-matrix pole (v') given by $\delta^{(1)}_4$ has a corresponding S-matrix pole (e') of the IS mode, forming a sharp
IS mode strength function peak which seems to be embedded in the background of the non-resonant continuum state
produced by $\delta^{(1)}_5$.
The $\delta^{(1)}_4$ component of the IS strength function also exhibits a very asymmetric shape. It is possible that
the Fano effect exists in this case as well, and if it exists, it needs to be clarified in future studies, including
what specific phenomena it may be related to. 

\begin{figure}[htbp]
\includegraphics[width=\linewidth]{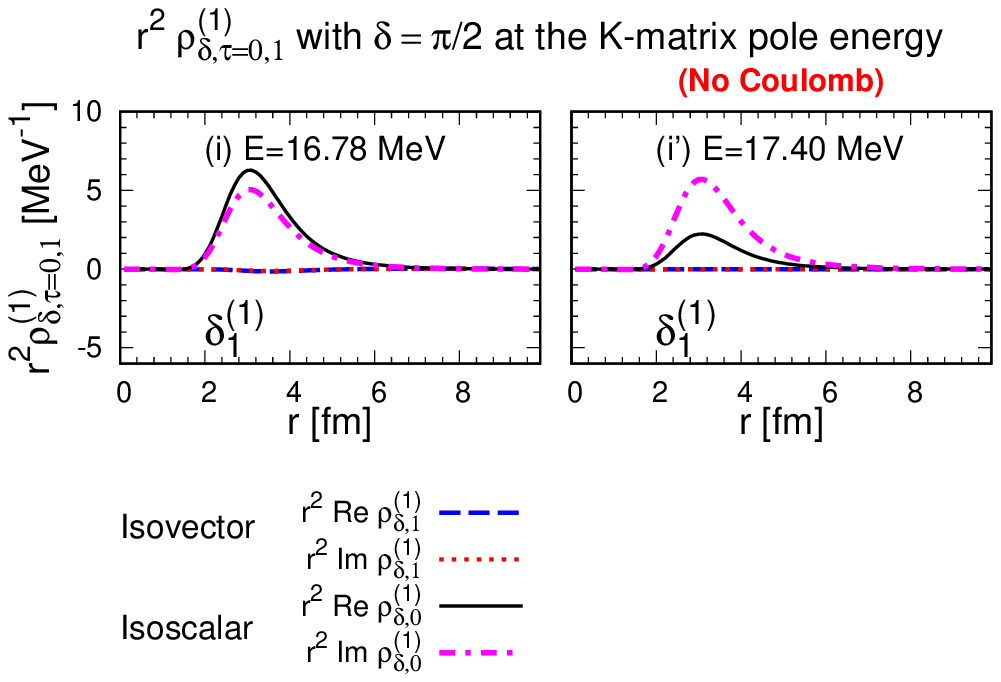}
\caption{(Color online)
The isovector transition density $\rho_{\delta,\tau=1}^{(1)}$ and
  isoscalar transition density $\rho_{\delta,\tau=0}^{(1)}$ at
  the first K-matrix pole energy of the eigenphase shift giving the
  K-matrix poles ((i) and (i')).}
\label{trdlow}
\end{figure}
\begin{figure}[htbp]
\includegraphics[width=\linewidth]{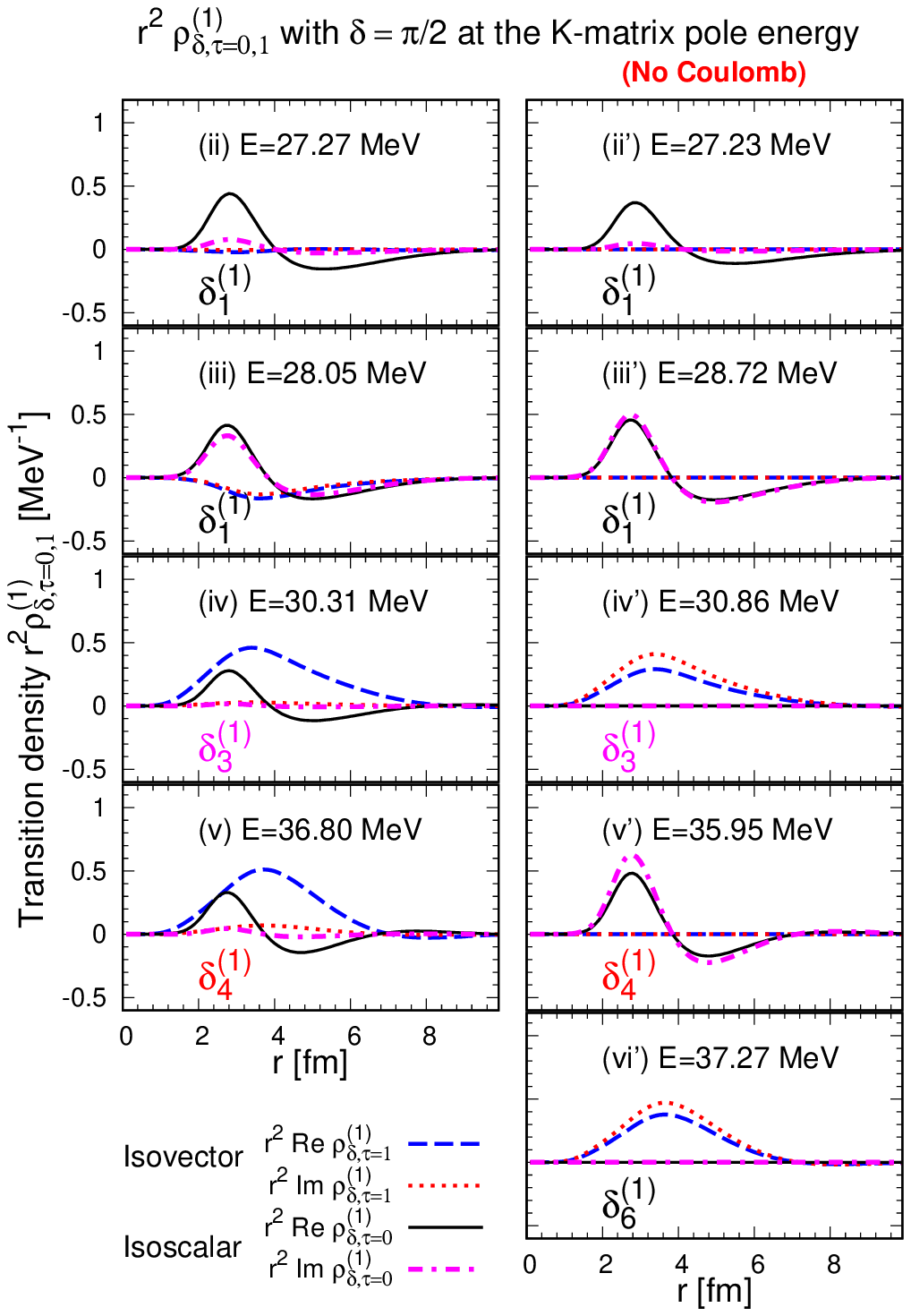}
\caption{(Color online)
  The isovector transition density $\rho_{\delta,\tau=1}^{(1)}$ and
  isoscalar transition density $\rho_{\delta,\tau=0}^{(1)}$ at
  the K-matrix pole energy of the eigenphase shift giving the
  K-matrix poles ((ii)-(v) and (ii')-(vi')).}
\label{trdGR}
\end{figure}

The transition density of the eigenphase shift component can be calculated using Eq. (\ref{trd-S1}).
Figs.\ref{trdlow} and \ref{trdGR} show the transition densities of the eigenphase shift components
which give their K-matrix poles at the K-matrix pole energy. 
The transition densities when neglecting the Coulomb interaction associated with Figs.\ref{srtphsO16L2noCLM}
and \ref{srtphsO16L2noCLMlow} are shown in the right-hand panels of Figs.\ref{trdlow} and \ref{trdGR}.
Note that the absolute square of the eigenphase shift component of the transition density integrated
with respect to $r$ gives the eigenphase shift component of the strength function in Figs.\ref{srtphsO16L2}
and \ref{srtphsO16L2noCLM}, and the sum of all the eigenphase shift components gives the total strength function,
by definition (Eq.(\ref{Sf-tdr-S1})).

The transition densities of the $\delta^{(1)}_1$ component of the K-matrix pole (i') in Fig. \ref{trdlow},
the $\delta^{(1)}_3$ component of (iv') and the $\delta^{(1)}_6$ component of (vi') in Fig. \ref{trdGR}
does not have nodes and has a peak near or slightly outside the surface of the nucleus. This shows that
the density distribution of neutrons and protons in the nucleus is collectively vibrating, which is
one of the typical properties of so-called collective vibration modes.

Since $\delta^{(1)}_1$ component of (i') has only the IS transition density amplitude and
$\delta^{(1)}_3$ component of (iv') and $\delta^{(1)}_6$ component of (vi') only the IV transition
density amplitude, respectively, this means that (i') is an IS collective vibration mode and (iv')
and (vi') are IV collective vibration modes. 

The $\delta^{(1)}_1$ component of the transition density at the K-matrix pole (ii') is the transition
density with an independent K-matrix pole, the $\delta^{(1)}_1$ component of (iii') and the $\delta^{(1)}_4$
component of (v') with corresponding S-matrix poles (c') and (e') respectively.
All these transition densities give contributions to the IS mode strength function only. 
Despite the $\delta^{(1)}_1$ component of (iii') and the $\delta^{(1)}_4$ component of (v') having
corresponding S-matrix poles (c') and (e') respectively, the transition densities have nodes and do not
show clear characteristics of collective vibration modes. 
In particular, even though the $\delta^{(1)}_4$ component of (v') gives a sharp peak in the shape of
the strength function of the IS mode, indicating the possible existence of a clear resonance, the
behavior of the transition density does not show any characteristic of collective vibration modes. 
This could be due to the possibility that resonance and collective mode do not necessarily coincide,
or it could be caused by special quantum interference effects such as the Fano effect.
This may need to be clarified in future research.

\begin{figure}[htbp]
\includegraphics[width=\linewidth]{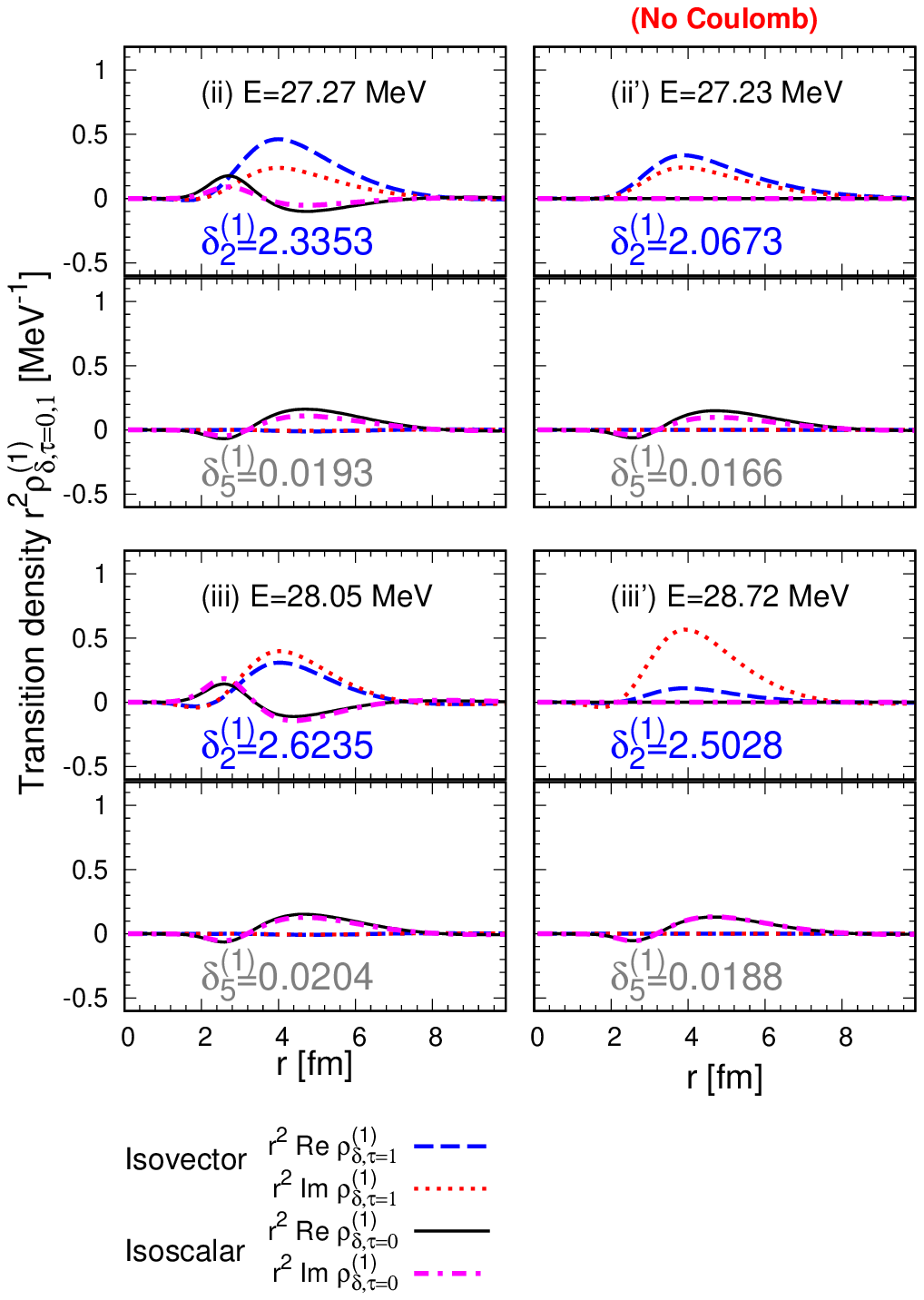}
\caption{(Color online)
  The $\delta^{(1)}_2$ and $\delta^{(1)}_5$ components of the transition densities
  (IS and IV modes) at K-matrix poles (ii), (iii), (ii') and (iii') respectively.}
\label{trdGR_ii-iii}
\end{figure}

Although $\delta^{(1)}_2$ and $\delta^{(1)}_5$ are eigenphase shifts which do not give K-matrix poles,
they make the main contribution to the IV and IS strength functions near the energies of the K-matrix
poles (ii') and (iii'), respectively. The transition densities of $\delta^{(1)}_2$ and $\delta^{(1)}_5$
at energies of the K-matrix poles (ii') and (iii') are shown in the right-hand panels of
Fig. \ref{trdGR_ii-iii}. 
The eigenphase shift $\delta^{(1)}_2$ corresponds to the corresponding S-matrix pole (b'), although it
does not have a K-matrix pole, and gives the main contribution to the first peak of the IV strength
function around 27 MeV, whose transition density exhibits the characteristics of the IV type collective
vibration.
The eigenphase shift $\delta^{(1)}_5$ has neither K-matrix poles nor S-matrix poles but makes the main
contribution to the IS strength function above $20$ MeV, and its transition density shows typical
non-resonant continuum state properties with nodes and widely spread amplitudes outside the nucleus.

Based on the knowledge gained from the analysis of calculations ignoring the Coulomb interaction that
has been carried out so far, the analysis of the RPA strength function including the Coulomb interaction
will be carried out from here onwards.
The difference caused by the presence or absence of the Coulomb interaction is the change in the
single-particle level structure of neutrons and protons due to isospin symmetry breaking
and the associated change in density distribution, and the mixing of IS and IV modes caused
by it. Comparing the strength functions in Figs.\ref{srtphsO16L2} and \ref{srtphsO16L2noCLM},
it can be seen that the IS and IV strength functions each contain a mixture of eigenphase shift
components which were not present when the Coulomb interaction was ignored.
There are also significant changes in the peak structure of the strength function associated
with changes in the eigenphase shift.

When the Coulomb interaction is ignored, the sharp peak of the IS strength function, which was formed
around $E = 36$ MeV with the contribution of the eigenphase shift $\delta^{(1)}_4$,
is split into two peaks.
Focusing on the change in the eigenphase shift $\delta^{(1)}_4$ shown in the bottom panels of
Figs.\ref{srtphsO16L2} and \ref{srtphsO16L2noCLM}, $\delta^{(1)}_4$ couples with $\delta^{(1)}_3$
and $\delta^{(1)}_6$ and the K-matrix pole (v') that existed below the $E=36.17$ MeV threshold
disappears.
A new K-matrix pole (v) is then formed near the K-matrix pole (vi') which was formed by
$\delta^{(1)}_6$ when the Coulomb interaction was ignored.
The eigenphase shift $\delta^{(1)}_4$ then gives contributions to both IS and IV strength. In fact, the
transition density of $\delta^{(1)}_4$ at the K-matrix pole (v), shown on the left-hand side of
Fig. \ref{trdGR}, now has contributions to both IS and IV modes.
The peak around $E=35$ MeV of the IS strength function is then formed by the independent
S-matrix pole (e), which no longer has a corresponding K-matrix pole.
A new peak around $E = 35$ MeV also appears in the IV strength function, which is formed by the
independent S-matrix pole (e) and is caused by the coupling of the eigenphase shift
$\delta^{(1)}_3$ with $\delta^{(1)}_4$.
The fact that the eigenphase shift $\delta^{(1)}_3$ is coupled to $\delta^{(1)}_4$ can be seen
from the transition densities.
The transition density of $\delta^{(1)}_3$ in the K-matrix pole (iv) shown in the left-hand
panel of Fig. \ref{trdGR} now includes characteristics of the transition density of
$\delta^{(1)}_4$ in (v') in the right-hand panel, which gives contributions to both IS
and IV strength.

The new peak around $E=28$ MeV in the IV strength function is caused by the contribution of
the eigenphase shift $\delta^{(1)}_1$. This is because $\delta^{(1)}_1$ couples with $\delta^{(1)}_2$,
so that $\delta^{(1)}_1$ now gives a contribution to the IV strength.
Conversely, by coupling $\delta^{(1)}_2$ with $\delta^{(1)}_1$, $\delta^{(1)}_2$ also makes the 
contribution to the IS strength when there is the Coulomb interaction, whereas originally it only
made the contribution to the IV strength when there was no Coulomb interaction.
The same can be seen from the $\delta^{(1)}_1$ and $\delta^{(1)}_2$ transition densities shown in
Figs. \ref{trdGR} and \ref{trdGR_ii-iii}. 
The effect of the Coulomb interaction on the eigenphase shift $\delta^{(1)}_5$ component, which is the main
component of the IS strength in the energy region above $20$ MeV, is very small. It can be seen
from the $\delta^{(1)}_5$ transition density shown in Fig. \ref{trdGR_ii-iii} that it hardly changes
at all.

The effect of the Coulomb interaction on the IS strength created by the eigenphase shift $\delta^{(1)}_1$,
which appears as a large and sharp peak at the lowest energy (K-matrix poles (i) and (i')) shown
in Figs.\ref{srtphsO16L2noCLMlow} and \ref{srtphsO16L2low}, is mainly in peak energy and width.
As far as Fig.\ref{srtphsO16L2low} is concerned, there is a slight mixing of the IV mode associated
with the coupling with $\delta^{(1)}_1$, but the mixing effect with the IV mode is small because
it is isolated away from the other poles.
The transition density (Fig.\ref{trdlow}) shows no effect of the mixing of IV modes on this pole,
only the characteristics of the IS collective vibration mode. 

\section{Summary and conclusion}
In this paper, we first derived the S-matrix which satisfies the unitarity using the Jost function
extended within RPA theory.
Considering the properties required to derive the RPA Green's function in Ref.\cite{JostRPA} and the symmetric
properties of the RPA Green's function and wave functions in complex energy space in the definition
of the RPA strength function, we were able to obtain the ``definition of the S-matrix with Jost function''
(Eqs. (\ref{S11def}) and (\ref{Sccdef})) which satisfies the unitarity and its associated
``scattering wave function'' (Eq. (\ref{psic})).
The S-matrix which satisfies the unitarity can be diagonalised by using the unitary matrices, and the diagonal
components of the diagonalised S-matrix are expressed in terms of eigenphase shifts. 
However, the original S-matrix gives a scattering boundary condition where the scattering
wavefunction leads through the S-matrix to a free particle state at $r\to\infty$ (the boundary
condition imposed to obtain irregular solutions).
Since RPA theory describes the excited states of the system in terms of the superposition of particle-hole
configurations defined by the mean-field caused by the effects of the residual interaction, Eq.(\ref{S1def}) was
adopted as the definition of the S-matrix from the idea of the two-potential problem to give the
eigenphase shift meaning to describe the RPA excited states.

Applying this method to the quadrupole excitation of the $^{16}$O, the strength function was then decomposed
into components for each eigenphase shift, and the correspondence between the eigenphase shift components
of the strength function and the S-matrix poles and eigenphase shifts was analyzed. 
The results show that the components of the RPA strength function corresponding to the eigenphase shift obtained
by diagonalising the S-matrix given by Eq.(\ref{S1def}) correspond to the S-matrix poles found on the complex energy
plane as the zeros of the determinant of the Jost function. 
Even though the peaks of the strength function correspond to S-matrix poles, there are not necessarily corresponding
K-matrix poles, and there are also independent S-matrix poles that do not have K-matrix poles.
Also, in the case of poles with simultaneous S- and K-matrix poles, if the eigenphase shift component of the strength
function corresponding to the pole exhibits an asymmetric shape, such that the presence of special interference effects
such as Fano resonances is suspected, the transition density of that component does not have the characteristics of
a collective mode.

The Fano effect is a special quantum interference effect which occurs when a bound or resonant state is coupled to a
continuum, and is known in atomic physics as the effect responsible for the phenomenon of autoionisation.
In Ref.\cite{jost-class}, it is shown that in nucleon-nucleus scattering, even in a resonant state where the S- and K-matrix
poles exist simultaneously, the scattering wavefunction loses its resonant feature (i.e. the feature that the scattering
amplitude is enhanced inside the nucleus) due to the presence of the Fano effect.
If a mechanism such as the Fano effect exists in RPA excited states, it is possible that the transition density loses
its collective mode character due to this effect. It remains possible, however, that the (resonance) states defined by
the S- and K-matrix poles are an independent concept and not directly related to the RPA collective mode.
In order to clarify this point, the further study is needed.

It is known that the solutions of the IS and IV modes can be obtained independently in RPA theory for $Z=N$ nuclei
such as $^{16}$O if the Coulomb interaction are neglected, and it is confirmed that the Jost function, its determinant
and the S-matrix derived in this paper can also be divided into IS and IV modes respectively.
Comparing calculations with and without the Coulomb interaction, it can be seen that they cause various effects, such
as the coupling of the eigenphase shifts of the IS and IV modes, the disappearance of the K-matrix poles, and the
creation of new peaks in the strength function and the splitting of the originally existing peak due to these effects.
The eigenphase shift component of the transition density corresponding to a given S-matrix pole is found to give
contributions to both IS and IV modes, as the S-matrix pole can no longer be divided into IS and IV modes when the
Coulomb interaction is present.

Judging comprehensively from these analyses, the eigenphase shifts obtained by diagonalising
the S-matrix derived in this paper can be considered to represent the ``RPA excited states''
corresponding to the S-matrix poles. 

\section*{Acknowledgments}

This work was supported by Hue University under the Core Research Program, Grant No. NCM.DHH.2018.09.
T.D.T. was funded by the Master, PhD Scholarship Programme of Vingroup Innovation Foundation (VINIF),
code no. VINIF.2023.TS127. 


\end{document}